\documentclass[11pt,a4paper]{article}

\usepackage[margin=1in]{geometry}
\usepackage[T1]{fontenc}
\usepackage[utf8]{inputenc}
\usepackage{lmodern}
\usepackage{microtype}

\usepackage{graphicx}
\usepackage{booktabs}
\usepackage{multirow}
\usepackage{xcolor}
\usepackage{amsmath}
\usepackage{amssymb}
\usepackage{siunitx}
\usepackage{subcaption}
\usepackage{threeparttable}
\usepackage{tikz}
\usetikzlibrary{shapes, arrows.meta, positioning, calc, patterns, fit}
\usepackage{pgfplots}
\pgfplotsset{compat=1.18}
\usepackage{placeins}
\usepackage{authblk}
\usepackage[colorlinks=true,citecolor=blue,urlcolor=blue,linkcolor=blue]{hyperref}
\usepackage[numbers,sort&compress]{natbib}

\newcommand{\demonstrated}{\textsc{[Demonstrated]}}
\newcommand{\simulated}{\textsc{[Simulated]}}
\newcommand{\projected}{\textsc{[Projected]}}
\newcommand{\nr}{\textsc{NR}}

\providecommand{\Description}[2][]{}

\title{Memristor Technologies for Dynamic Vision Sensors:\\ A Critical Assessment and Research Roadmap}

\author[1]{Mohamad Yazan Sadoun}
\author[2]{Edris Zaman Farsa}
\author[1]{Sarah Sharif}
\author[1,*]{Yaser Mike Banad}

\affil[1]{University of Oklahoma, Norman, Oklahoma, USA}
\affil[2]{Institute of Computer Engineering, Technische Universit\"at Wien (TU~Wien), Vienna, Austria}
\affil[*]{Corresponding author: \texttt{bana@ou.edu}}

\date{}

\begin{document}

\maketitle

\begin{abstract}
Edge-AI deployment is bottlenecked by data-movement energy; pairing event-driven vision sensors with in-memory analog compute could lift that ceiling by orders of magnitude. Both technologies are individually mature; the framework distinguishing fabricated demonstrations from projected systems is missing. Of six application domains surveyed (robotics, autonomous vehicles, AR/VR, surveillance, medical imaging, IoT), half rest entirely on projection, and existing hardware sits at Technology Readiness Levels 2--5. This evidence-graded review applies a three-paradigm architectural taxonomy and benchmarks the gap against current digital neuromorphic alternatives. It identifies an end-to-end integrated DVS-memristor system as the field's open challenge, with testable accuracy and power targets.
\end{abstract}


\section{Introduction}
\label{sec:introduction}

Dynamic vision sensors (DVS) encode visual information as asynchronous streams of brightness-change events, each carrying a microsecond-resolution timestamp, pixel address, and polarity~\cite{lichtsteiner2008dvs}.
This event-driven paradigm eliminates the redundant transmission of static pixels that dominates frame-based imaging, producing data rates that scale with scene dynamics rather than sensor resolution.
Yet the downstream processing pipeline remains overwhelmingly digital: events are buffered in DRAM, shuttled across a memory bus, and processed by clock-driven arithmetic units.
The resulting von Neumann bottleneck (repeated data movement between spatially separated memory and compute) consumes orders of magnitude more energy than the sensor itself and erodes the latency and power advantages that event-driven sensing was designed to deliver~\cite{ielmini2018memory, sebastian2020memory}.
Memristive devices offer a direct remedy.
By co-locating nonvolatile storage and analog multiplication within a single two-terminal element, memristor crossbar arrays can execute multiply-accumulate operations in place, eliminating the dominant data-movement cost~\cite{strukov2008missing, prezioso2015training}.
Pairing an inherently sparse, event-driven sensor front-end with an in-memory computing back-end is therefore a structurally natural match, and one that several recent device demonstrations have begun to explore~\cite{dang2024multiphototransistor, yoon2024memristor}.

Several reviews have surveyed adjacent portions of this design space without treating the DVS--memristor intersection jointly. Event-vision surveys cover algorithms, applications, and methods (Gallego et al.~\cite{gallego2022event}; Chakravarthi et al.~\cite{chakravarthi2024event}; Cazzato and Bono~\cite{cazzato2024application}) but contain no memristive or analog-compute hardware. Device-side reviews map optoelectronic memristors and in-sensor computing (Wang et al.~\cite{wang2024roadmap}; Cho et al.~\cite{cho2022progress}; Wan et al.~\cite{wan2023insensor}; Su et al.~\cite{su2024nonvon}; Zhang et al.~\cite{chen2023emerging}) but treat DVS integration only peripherally. Five foundational neuromorphic surveys~\cite{kendall2020building, markovic2020physics, christensen2022roadmap, schuman2022neuromorphic, kudithipudi2025neuromorphicscale} establish the field's vocabulary but do not address the memristor--DVS intersection.

Ten recent reviews narrow the gap. Mao et al.\ (2026)~\cite{mao2026photoelectric} treat photoelectric memristors for visual perception in depth but do not classify integration paradigms as distinct architectural tiers. Three 2025 surveys cover neuromorphic visual systems and optoelectronic memristors~\cite{liu2025neuromorphic, pan2025optoelectronic, ren2025optoelectronic} without providing a DVS--memristor taxonomy or evidence grading. Shooshtari et al.~\cite{shooshtari2025review_imc_snn} cover memristors for IMC/SNN but not DVS. Five recent event-vision surveys~\cite{cimarelli2025dvs_review, zheng2024event_survey, aliakbarpour2024dvssurvey, iaboni2024snn_review, moitra2024snn_imc} cover event-vision hardware/algorithms or SNN co-design but either contain no memristor content or do not grade claims by evidence. Table~\ref{tab:review_comparison} summarizes coverage against these prior works across nine thematic dimensions.\footnote{Assessment reflects depth of treatment, not mere mention. \checkmark\ = comprehensive, dedicated section(s) with critical analysis; $\circ$ = partial, addressed but not the primary focus; $\times$ = not covered.}

\begin{table*}[t]
\centering
\caption{Comparison of Coverage Dimensions Across Existing Reviews. \checkmark\ = Comprehensive coverage with quantitative analysis; $\circ$ = Partial coverage or brief mention; $\times$ = Not covered. Assessment reflects depth of treatment, not mere mention.}
\label{tab:review_comparison}
\resizebox{\textwidth}{!}{%
\begin{threeparttable}
\footnotesize
\setlength{\tabcolsep}{3pt}
\begin{tabular}{@{}l c c c c c c c c c@{}}
\toprule
\textbf{Review} & \rotatebox{70}{\textbf{DVS Fundamentals}} & \rotatebox{70}{\textbf{Memristor Physics}} & \rotatebox{70}{\textbf{In-Sensor Computing}} & \rotatebox{70}{\textbf{Crossbar Architectures}} & \rotatebox{70}{\textbf{Material Systems}} & \rotatebox{70}{\textbf{CMOS Integration}} & \rotatebox{70}{\textbf{Application Analysis}} & \rotatebox{70}{\textbf{Co-Design Strategies}} & \rotatebox{70}{\textbf{Research Roadmap}} \\
\midrule
Gallego et al. (2022) \cite{gallego2022event}       & \checkmark & $\times$ & $\times$ & $\times$ & $\times$ & $\times$ & \checkmark & $\times$ & $\circ$ \\
Cho et al. (2022) \cite{cho2022progress}             & $\times$ & $\circ$ & \checkmark & $\times$ & \checkmark & $\times$ & \checkmark & $\times$ & $\circ$ \\
Wan et al. (2023) \cite{wan2023insensor}             & $\times$ & $\circ$ & \checkmark & $\times$ & \checkmark & $\circ$ & $\circ$ & $\times$ & \checkmark \\
Chakravarthi et al. (2024) \cite{chakravarthi2024event} & \checkmark & $\times$ & $\times$ & $\times$ & $\times$ & $\times$ & \checkmark & $\times$ & $\circ$ \\
Cazzato \& Bono (2024) \cite{cazzato2024application} & \checkmark & $\times$ & $\times$ & $\times$ & $\times$ & $\times$ & \checkmark & $\times$ & $\times$ \\
Wang Z. et al. (2024) \cite{wang2024roadmap}         & $\times$ & \checkmark & \checkmark & $\circ$ & \checkmark & $\circ$ & $\circ$ & $\times$ & \checkmark \\
Wang H. et al. (2024) \cite{su2024nonvon}            & $\circ$ & $\circ$ & \checkmark & $\circ$ & \checkmark & $\circ$ & $\circ$ & $\times$ & $\circ$ \\
Zhang et al. (2023) \cite{chen2023emerging}          & $\times$ & \checkmark & $\circ$ & $\times$ & \checkmark & $\times$ & $\circ$ & $\times$ & $\circ$ \\
\midrule
\multicolumn{10}{@{}l}{\textit{Recent competitor reviews (2024--2026):}} \\
Mao et al. (2026) \cite{mao2026photoelectric}        & $\times$ & \checkmark & \checkmark & $\times$ & \checkmark & $\times$ & $\circ$ & $\times$ & \checkmark \\
Liu et al. (2025) \cite{liu2025neuromorphic}         & $\times$ & \checkmark & \checkmark & $\circ$ & \checkmark & $\times$ & \checkmark & $\times$ & $\circ$ \\
Pan et al. (2025) \cite{pan2025optoelectronic}       & $\times$ & \checkmark & \checkmark & $\times$ & \checkmark & $\times$ & $\circ$ & $\times$ & $\circ$ \\
Ren et al. (2025) \cite{ren2025optoelectronic}       & $\times$ & $\circ$ & \checkmark & $\circ$ & \checkmark & $\circ$ & \checkmark & $\times$ & $\circ$ \\
Shooshtari et al. (2025) \cite{shooshtari2025review_imc_snn} & $\times$ & \checkmark & $\circ$ & \checkmark & \checkmark & $\circ$ & $\circ$ & $\times$ & $\circ$ \\
Cimarelli et al. (2025) \cite{cimarelli2025dvs_review} & \checkmark & $\times$ & $\times$ & $\times$ & $\times$ & $\times$ & \checkmark & $\times$ & $\circ$ \\
Iaboni et al. (2024) \cite{iaboni2024snn_review} & \checkmark & $\times$ & $\times$ & $\times$ & $\times$ & $\times$ & $\circ$ & $\times$ & $\circ$ \\
Zheng et al. (2024) \cite{zheng2024event_survey} & \checkmark & $\times$ & $\times$ & $\times$ & $\times$ & $\times$ & \checkmark & $\times$ & $\circ$ \\
Aliakbarpour et al. (2024) \cite{aliakbarpour2024dvssurvey} & \checkmark & $\times$ & $\times$ & $\times$ & $\times$ & $\times$ & \checkmark & $\times$ & $\circ$ \\
Moitra et al. (2024) \cite{moitra2024snn_imc} & $\times$ & \checkmark & $\circ$ & \checkmark & \checkmark & $\circ$ & $\circ$ & $\circ$ & $\circ$ \\
\midrule
\textbf{This review}                                  & \checkmark & \checkmark & \checkmark & \checkmark & \checkmark & $\circ^\ddagger$ & \checkmark$^\dagger$ & $\circ$ & \checkmark \\
\bottomrule
\end{tabular}
\begin{tablenotes}
\small
\item $^\dagger$ Application analysis uses evidence-graded framework separating demonstrated, simulated, and projected capabilities.
\item $^\ddagger$ CMOS integration is addressed in a single subsection (§6.2) rather than as a dedicated deep dive, so it is marked $\circ$ rather than \checkmark despite being covered.
\item Co-Design Strategies is marked $\circ$ rather than \checkmark: this review identifies co-design as essential but does not propose a complete co-design methodology.
\item The novelty claim in Section~\ref{sec:introduction}---that contributions C1 (R1/R2/R3 architectural taxonomy), C2 (\demonstrated/\simulated/\projected{} evidence-graded claim framework), C3 (double baseline against memristor-CIM SOTA \emph{and} shipping digital neuromorphic processors), and C4 (falsifiable quantitative roadmap with N1/M1/L1 milestones) are not jointly applied by prior surveys---is supported by the table as follows: no row above ``This review'' carries \checkmark{} on \emph{both} Crossbar Architectures and Application Analysis with an explicit evidence-grading column or a digital-neuromorphic baseline column; the closest competitors (Mao~2026, Liu~2025, Moitra~2024) cover one or two of the four instruments individually but not the joint set. The table dimensions intentionally describe topical \emph{coverage}; the four-instrument novelty argument operates on \emph{methodological} axes not enumerated as columns.
\end{tablenotes}
\end{threeparttable}%
}
\end{table*}

\paragraph{Why a critical assessment of this intersection now.}
Both adjacent fields are individually mature: commercial DVS sensors operate at TRL~8--9 with sub-5~$\mu$m pixels and megapixel resolutions~\cite{finateu2020gen4, suh2020samsung}, while memristor compute-in-memory has reached industrial deployment in fabricated 256$\times$256 crossbars~\cite{rao2023thousands}, 48-core RRAM-CIM chips~\cite{wan2022neurram}, fully hardware-implemented memristor CNNs~\cite{yao2020fully}, equivalent-accuracy training with analog memory~\cite{ambrogio2018equivalent}, high-precision programmable arrays~\cite{song2024precision}, and PCM-based analog AI processors~\cite{ambrogio2023analogai, legallo2023mixedsignal}. Ten adjacent reviews published in 2024--2026~\cite{mao2026photoelectric, liu2025neuromorphic, pan2025optoelectronic, ren2025optoelectronic, shooshtari2025review_imc_snn, cimarelli2025dvs_review, zheng2024event_survey, aliakbarpour2024dvssurvey, iaboni2024snn_review, moitra2024snn_imc} establish that the integration is now seen as imminent across both communities. What is missing is the framework that distinguishes fabricated demonstrations from projected systems and turns the integration target into a measurable engineering goal. The four instruments this review contributes (taxonomy, evidence grading, double baseline, falsifiable roadmap) supply that framework: each integration claim becomes a testable hypothesis with a concrete target, and the roadmap converts the open opportunity into milestones the community can pursue.

This review is organized around three research questions that target the gaps identified above.
\textbf{RQ1:} What architectural paradigms have been proposed or demonstrated for integrating memristive devices with event-based vision sensors, and what are their comparative performance trade-offs?
\textbf{RQ2:} What material systems and device technologies enable memristor--DVS integration, and what are the device-level barriers (variability, endurance, CMOS compatibility) to scaling?
\textbf{RQ3:} What is the current maturity of memristor--DVS systems across application domains, and what specific gaps remain between laboratory demonstrations and practical deployment?

\paragraph{Scope and audience.}
The intersection's small fabricated footprint shapes how this review reads. Of 146 retained publications, ten operate at the device-level memristor--DVS boundary, and seven couple a fabricated memristor to event-driven or light-driven input (Section~\ref{sec:methodology_limitations}). This work is therefore framed as a \emph{structured critical assessment with a roadmap}: aimed at researchers who need to separate fabricated results from simulation and projection, and at programme managers who need a falsifiable milestone set for funding decisions. Readers seeking a primer on event-vision algorithms or memristor physics in isolation are better served by Gallego et al.~\cite{gallego2022event} and Wang et al.~\cite{wang2024roadmap} respectively; this review's contribution is the joint critical assessment those single-field surveys do not provide.

The review covers three distinct roles that memristive devices play in event-driven vision systems: (1)~the \emph{photomemristor as sensor}, where the memristive element itself transduces light and encodes temporal contrast; (2)~\emph{in-pixel memory}, where a memristor embedded within the pixel circuit stores per-pixel state such as background models or adaptation thresholds; and (3)~the \emph{crossbar accelerator}, where off-pixel memristor arrays perform inference on accumulated event representations.

Four analytical instruments distinguish this assessment from the prior surveys in Table~\ref{tab:review_comparison}, which apply some of these tools individually but not in combination.
\textbf{(C1)}~A three-paradigm architectural taxonomy (R1 photomemristor / R2 in-pixel memory / R3 crossbar accelerator) classifying every intersection work by where the memristor sits in the signal path, cross-indexed against a seven-material, six-requirement compatibility matrix.
\textbf{(C2)}~An evidence-graded claim framework (\demonstrated, \simulated, \projected) applied to every application claim, revealing that three of six commonly cited application domains rest entirely on simulation or projection.
\textbf{(C3)}~A double baseline against current memristor-CIM SOTA~\cite{wan2022neurram, rao2023thousands, ambrogio2023analogai, legallo2023mixedsignal} and shipping digital neuromorphic processors (Loihi~2, SynSense Speck, SpiNNaker~2), making the gap falsifiable against two moving targets prior reviews omit.
\textbf{(C4)}~A falsifiable research roadmap whose near-term milestone N1 ($\geq$90\% DVS128-Gesture at $\leq$10\,mW, $\leq$5\,ms) and mid-term M1 ($\geq$640$\times$480 DVS+memristor at $\geq$85\%, $\leq$50\,mW) are provable or disprovable, not aspirational.

The remainder of the paper unfolds as a single argument rather than as independent survey chapters. Section~\ref{sec:methodology} sets the screening protocol that defines the corpus. Section~\ref{sec:foundations} extracts the device-level and sensor-level numbers that every later claim is tested against, and closes with the explicit case against memristors (Section~\ref{sec:case-against}). Section~\ref{sec:architectures} organises the fabricated literature into the R1/R2/R3 taxonomy and exposes the cross-cutting patterns that the device-level numbers predict. Section~\ref{sec:applications} applies the evidence-graded framework to every application claim in the corpus, producing the maturity assessment of the field. Section~\ref{sec:challenges} converts the patterns and gaps into concrete technical barriers and a falsifiable roadmap. Section~\ref{sec:conclusion} closes by answering the three research questions against the corpus.

\section{Structured Review Methodology}
\label{sec:methodology}

This is a structured critical narrative review of the memristor--DVS intersection, using the PRISMA reporting guidelines~\cite{page2021prisma} for transparency. The analytical contributions (architectural taxonomy, evidence-graded framework, falsifiable roadmap) are critical-narrative instruments designed for a nascent field where standardised terminology has not yet converged across device physics, circuit design, and computer vision.

Boolean queries combining memristor-family terms (memristor, RRAM, ReRAM, resistive switching, in-memory computing, crossbar) with event-vision terms (DVS, event camera, silicon retina, address-event representation) were submitted on 2026-05-01 to IEEE Xplore, Web of Science, and OpenAlex, scoped to title plus abstract plus indexed-keyword fields. Forward and backward citation chaining was applied to ten core intersection works~\cite{vourkas2021inmemory, laiho2015memristive, cao2023dual, cai2023insitu, wang2024perovskite, huang2025multimode, tan2023dynamic, li2018analogue, yoon2024memristor, zhou2023silk} and ten adjacent reviews to ensure coverage of papers using non-standard keyword vocabulary. The temporal scope spanned January 2008 (the year of both the first DVS JSSC publication~\cite{lichtsteiner2008dvs} and the HP Labs memristor demonstration~\cite{strukov2008missing}) through April 2026. Patents, conference abstracts under two pages, and non-English publications were excluded \emph{a priori}.

\begin{figure*}[!htbp]
\centering
\includegraphics[width=\textwidth]{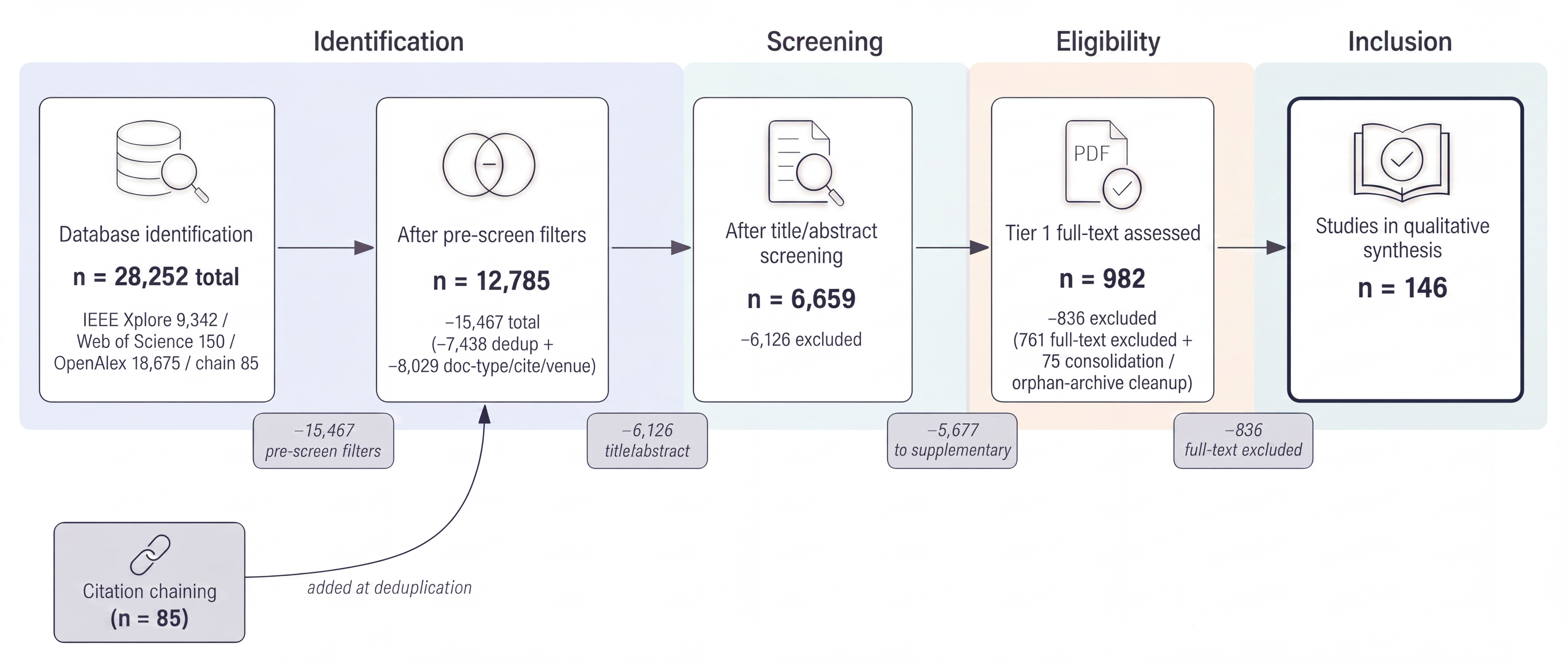}
\Description{Horizontal PRISMA-style flow diagram showing the staged literature screening pipeline. From 28,252 records identified across IEEE Xplore, Web of Science, OpenAlex and citation chaining, deduplication removed 7,438 database duplicates leaving 20,814 records, document-type / citation-threshold / venue-tier filtering removed a further 8,029 to 12,785, title-and-abstract screening retained 6,659 eligible records, and full-text-equivalent screening of the 982 highest-priority records produced 146 papers in the final qualitative-synthesis corpus.}
\caption{Structured PRISMA-style screening pipeline using the PRISMA reporting guidelines~\cite{page2021prisma} for transparency. From 28{,}252 records identified across IEEE Xplore, Web of Science, OpenAlex and citation chaining, the staged reduction produced a final corpus of 146 publications spanning memristor device physics, DVS sensor engineering, and their intersection; ten papers operate at the direct intersection (the most complete chip-prototype reference design is detailed in Section~\ref{sec:array-level}), and approximately 50 adjacent works provide context.}
\label{fig:prisma}
\end{figure*}

Of 28{,}252 records identified, deduplication and document-type, citation-threshold, and venue-tier filtering reduced the pool to 12{,}785 records for screening; title-and-abstract assessment retained 6{,}659 eligible records; the 982 highest-priority records (ranked by composite relevance score combining citation-chain origin, log-citation impact, and recency) advanced to full-text-equivalent screening, producing the final corpus of 146 publications. Figure~\ref{fig:prisma} shows the staged reduction. The complete screening protocol, per-stage exclusion-reason breakdowns, search-string variants, and the machine-readable audit trail are documented in Supplement~S1.

\subsection{Scope of the Corpus}
\label{sec:methodology_limitations}

Three scope notes anchor the evidence grading that follows.

\textbf{Small true intersection.} Ten publications occupy the memristor--DVS conceptual boundary~\cite{vourkas2021inmemory, laiho2015memristive, cao2023dual, cai2023insitu, wang2024perovskite, huang2025multimode, tan2023dynamic, li2018analogue, yoon2024memristor, zhou2023silk}; seven of these strictly couple a fabricated device to event-driven or light-driven input~\cite{laiho2015memristive, cao2023dual, cai2023insitu, wang2024perovskite, huang2025multimode, tan2023dynamic, zhou2023silk}, with the remaining three being a circuit-simulation study~\cite{yoon2024memristor}, a SPICE-architectural analysis~\cite{vourkas2021inmemory}, and a generic 128$\times$64 HfO\textsubscript{2} crossbar with no DVS coupling reported~\cite{li2018analogue}. The other 143 papers in the corpus provide context (device physics, sensor engineering, neuromorphic architectures); approximately 50 of these carry the bulk of the cross-cutting argument, and the remainder appear as supporting-context references for specific numbers and benchmarks.

\textbf{Gray literature and preprints.} Including arXiv preprints broadens coverage of a fast-moving field but introduces work that has not undergone full peer review; preprint-only sources are flagged in the reference list where applicable.

\textbf{Post-cutoff additions.} Database queries were executed on 2026-05-01 with a temporal cutoff of 2026-04-30. A small number of high-impact papers entered the corpus through forward citation chaining of the core intersection works and through targeted manual addition of journal-of-record references; the supplementary corpus listing identifies each by its citation-chain provenance.

\section{Foundations}
\label{sec:foundations}

This section is not a free-standing primer; every device-level fact summarised here is recruited later as a test that a memristor-DVS claim must survive. Memristor switching speed, endurance, and variability set the bar for any in-pixel or in-array integration claim made in Section~\ref{sec:architectures}. DVS pixel pitch and event-rate budgets dictate which integration roles are physically possible. The ``competing technologies'' subsection is the warm-up for the critical case against memristors raised in Section~\ref{sec:case-against} and quantified throughout Sections~\ref{sec:applications}--\ref{sec:challenges}. Readers familiar with memristor physics or event-camera operation can skim the relevant subsection, but should not treat the device-level numbers as background --- the integration requirements derived in this section (variability, endurance, switching speed, retention, CMOS compatibility, optical sensitivity) anchor the falsifiable roadmap milestones in Section~\ref{sec:roadmap}.

\subsection{Memristor Device Fundamentals}
\label{sec:memristor-fundamentals}

\begin{figure*}[!htbp]
\centering
\includegraphics[width=\textwidth]{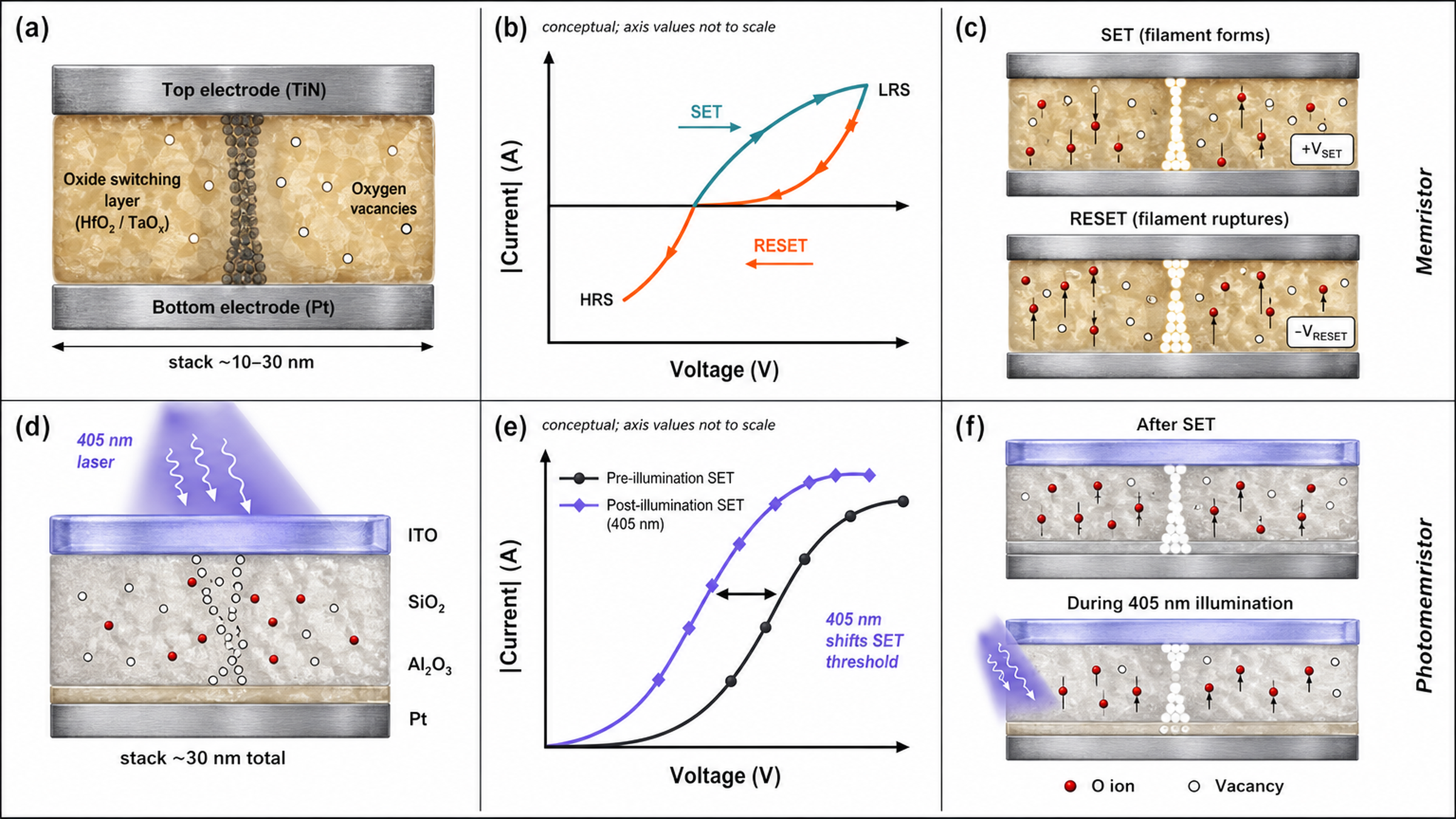}
\Description{Six-panel device-physics primer arranged as a 2x3 grid. TOP ROW (Memristor): (a) photo-realistic cross-section of a metal-insulator-metal memristor stack with TiN top electrode, HfO2/TaOx oxide switching layer containing oxygen vacancies, and Pt bottom electrode (10-30 nm total); (b) measured bipolar I-V hysteresis on log current scale showing SET and RESET branches characteristic of conductive-filament memristors; (c) atomistic switching mechanism in two stacked sub-frames showing filament formation during SET (positive bias) and rupture during RESET (negative bias). BOTTOM ROW (Photomemristor): (d) photo-realistic cross-section of an ITO/SiO2/Al2O3/Pt photomemristor stack illuminated by a 405 nm laser-light cone, showing oxygen ions (red dots) and vacancies (open circles) in the oxide layers (~30 nm total); (e) pre- and post-illumination SET current-voltage curves showing how 405 nm illumination shifts the SET threshold; (f) atomistic oxygen-vacancy schematic in two stacked sub-frames showing the post-SET filament and the photon-driven ion redistribution that breaks the filament near the top electrode.}
\caption{Memristor and photomemristor device physics. \textbf{Top row (Memristor):} (a)~photo-realistic cross-section of a metal--insulator--metal (MIM) memristor stack: TiN top electrode, HfO\textsubscript{2}/TaO\textsubscript{x} oxide switching layer with oxygen vacancies, and Pt bottom electrode (stack thickness 10--30\,nm); (b)~conceptual bipolar I--V hysteresis of a conductive-filament memristor showing the SET (rising) and RESET (falling) branches, with low- and high-resistance states (LRS/HRS) labelled --- the plot is illustrative; axis values are not to scale; (c)~atomistic switching mechanism: filament forms under positive bias (top sub-frame, SET) and ruptures under negative bias (bottom sub-frame, RESET). \textbf{Bottom row (Photomemristor):} (d)~ITO/SiO\textsubscript{2}/Al\textsubscript{2}O\textsubscript{3}/Pt stack under 405\,nm illumination, showing the oxygen-ion and vacancy distribution ($\sim$30\,nm total); (e)~conceptual pre- and post-illumination SET I--V characteristics under 405\,nm optical excitation, illustrating how the photon flux shifts the SET threshold --- axis values are not to scale; (f)~atomistic oxygen-vacancy model contrasting the post-SET filament (top sub-frame) with the photon-driven ion redistribution during illumination (bottom sub-frame), the mechanism underlying negative photoconductivity in this stack.}
\label{fig:memristor_switching}
\end{figure*}

\subsubsection{Device structure and switching mechanisms.}
Chua predicted the memristor in 1971 as the missing fourth fundamental circuit element linking charge and magnetic flux~\cite{chua1971memristor}. Strukov et al.\ provided the first experimental realization in 2008 using a thin TiO\textsubscript{2} film sandwiched between platinum electrodes---a metal--insulator--metal (MIM) stack whose resistance depends on the history of applied voltage~\cite{strukov2008missing}. Two principal switching mechanisms dominate current devices. \emph{Filamentary switching} forms and ruptures a conductive filament (typically oxygen vacancies or metal ions) through the insulating layer, producing abrupt resistance changes with high ON/OFF ratios but stochastic filament geometry. \emph{Interfacial (area-type) switching} modulates a Schottky barrier or tunnel junction across the entire electrode interface, yielding more gradual and uniform transitions at the cost of lower ON/OFF contrast~\cite{yang2013memristive,gooran2025electrode}. A third class --- \emph{electrochemical metallisation} or CBRAM, in which metal cations (Ag, Cu) plate and dissolve to form an atomic-scale filament --- is treated alongside competing technologies in Section~\ref{sec:why-memristors}. Recent device-level reviews provide complementary surveys of memristor synapse design~\cite{chandrasekaran2026synapse} and current opinions on memristor-accelerated machine-learning hardware~\cite{jiang2025opinions_memristor}.

\subsubsection{Performance metrics.}
Four device-level metrics govern suitability for vision applications. \emph{Switching speed} ranges from sub-10~ns in optimised HfO\textsubscript{x}/TaO\textsubscript{x} devices~\cite{yang2013memristive} to microseconds in organic films; matching DVS temporal resolution ($<$1~ms) requires speeds no slower than $\sim$10~\si{\micro\second}. \emph{Endurance} spans $10^{6}$--$10^{9}$ cycles for typical metal-oxide stacks, with advanced devices exceeding $10^{12}$~\cite{ielmini2018memory}; sustained DVS write rates can stress even high-endurance devices (Section~\ref{sec:challenges}). Throughout this review, $10^{6}$--$10^{9}$ refers to the routinely reported endurance range, $10^{12}$ to advanced-device demonstrations, and the $10^{10}$ figure used in Milestone~N2 (Section~\ref{sec:roadmap}) to the criterion required for sustained DVS-grade workloads. \emph{Retention} exceeds 10~years for well-optimized stacks but degrades at elevated temperatures, raising automotive reliability concerns~\cite{gooran2025thermal}. \emph{Device-to-device variability} remains the most stubborn obstacle: C2C and D2D resistance spread is typically 5--30\% (outliers: $<$1\% in ferroelectric stacks, $>$19\% in 2D arrays; see Section~\ref{sec:device-challenges}), directly corrupting the analog weight precision crossbar arrays require~\cite{ielmini2018memory}.

\subsubsection{Crossbar arrays and analog compute.}
Memristors arranged in a crossbar (rows as input lines, columns as output lines, a device at each junction) perform matrix--vector multiplication in a single step: input voltages $V_i$ applied to rows produce column currents $I_j = \sum_i G_{ij} V_i$ via Ohm's law and Kirchhoff's current law, eliminating the data-movement bottleneck of von Neumann architectures~\cite{ielmini2018memory}. Parasitic wire resistance, sneak-path currents in passive arrays, and ADC overhead limit practical array sizes and energy efficiency (Section~\ref{sec:challenges}). Compact models for circuit simulation include the HP linear drift model~\cite{strukov2008missing}, the threshold-based TEAM model~\cite{kvatinsky2013team}, and material-specific models for TaO\textsubscript{x} filament dynamics~\cite{gooran2024taox}. Figure~\ref{fig:memristor_switching}(d--f) extends the physics primer to the photomemristor case discussed in Section~\ref{sec:architectures}.

\subsection{Dynamic Vision Sensor Principles}
\label{sec:dvs-principles}

\begin{figure*}[!htbp]
\centering
\includegraphics[width=\textwidth]{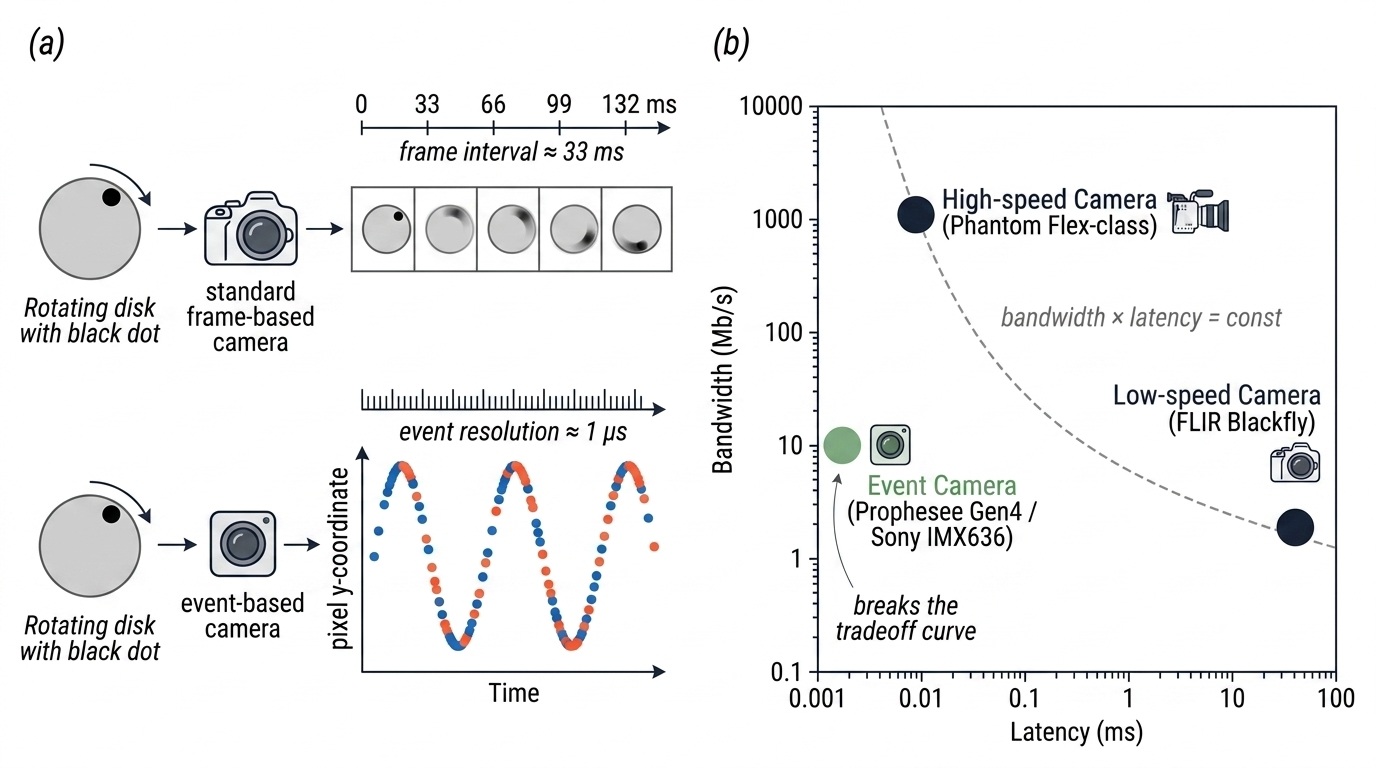}
\Description{Two-panel comparison. Panel (a): a rotating-disk scene viewed by a conventional frame camera (producing dense, motion-blurred frames) versus an event camera (producing sparse polarity events with microsecond timestamps). Panel (b): a log-log bandwidth-versus-latency plot showing frame cameras lying on the bandwidth/latency trade-off curve while event cameras achieve simultaneously low bandwidth and low latency.}
\caption{Comparison of frame-based and event-based camera outputs. (a)~A rotating-disk scene viewed by both sensors: a conventional camera captures full frames at fixed intervals, producing redundant data and motion blur; an event camera responds only to brightness changes, generating sparse, temporally precise event tuples $(x, y, t, p)$. (b)~Bandwidth--latency trade-off: conventional high- and low-speed cameras trade bandwidth for latency ($\text{bandwidth} \propto 1/\text{latency}$), whereas event cameras achieve low latency at low bandwidth simultaneously.}
\label{fig:frame_vs_event}
\end{figure*}

\subsubsection{Event-driven sensing paradigm.}
\begin{figure}[!htbp]
\centering
\includegraphics[width=\columnwidth]{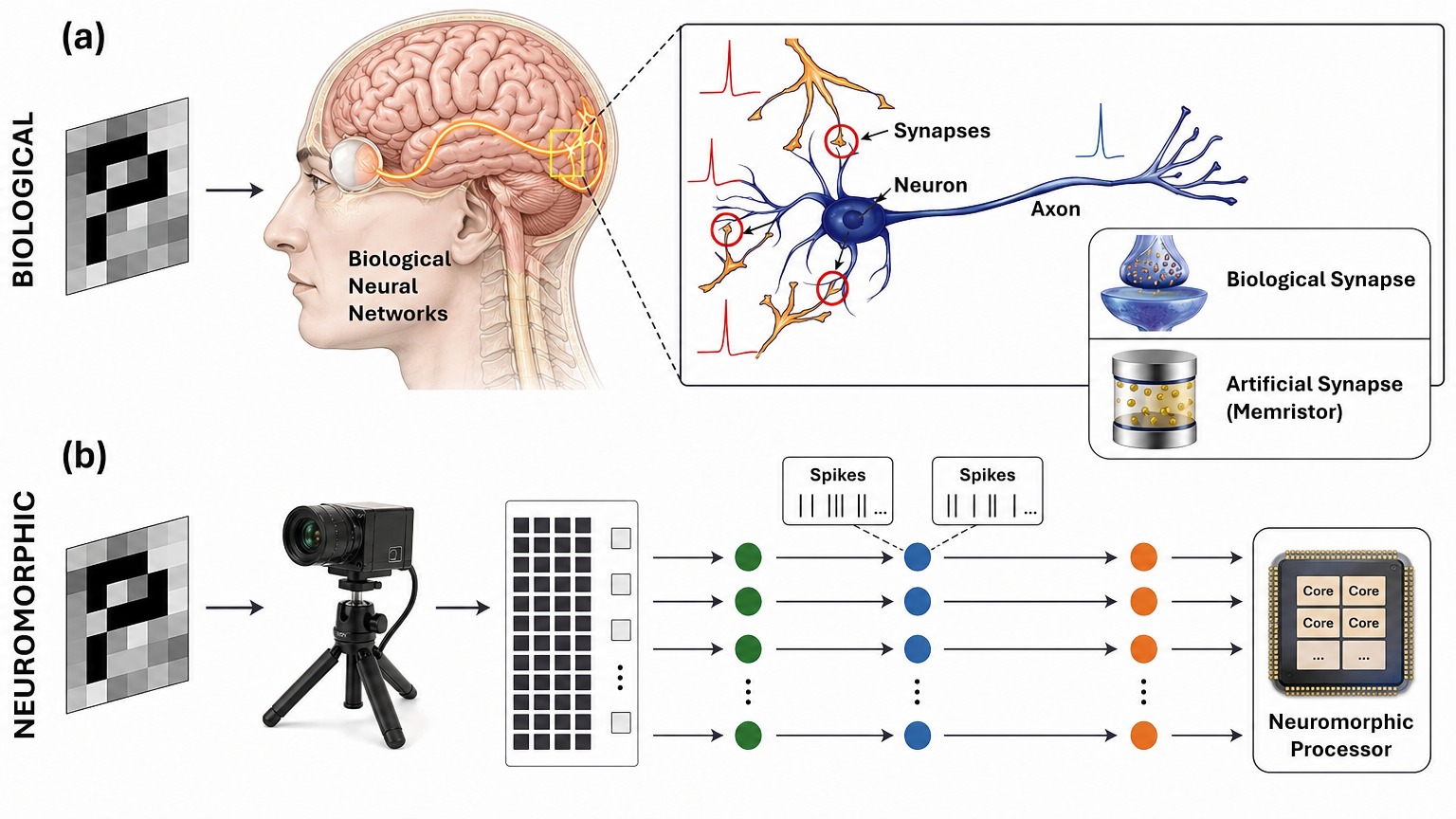}
\Description{Two-row comparative diagram. Top row: biological visual processing through synapses, neurons, and neural networks reaching the brain. Bottom row: neuromorphic emulation using memristive artificial synapses, spiking neuron circuits, crossbar arrays, and a neuromorphic processor.}
\caption{Biological versus neuromorphic vision systems. Top: biological neural networks process visual stimuli through synapses, neurons, and networks to the brain. Bottom: neuromorphic systems use memristive artificial synapses, spiking neurons, crossbar array architectures, and neuromorphic processors to emulate biological visual processing.}
\label{fig:bio_vs_neuromorphic}
\end{figure}

A dynamic vision sensor (DVS) replaces the synchronous global-shutter readout of conventional cameras with independent, asynchronous per-pixel operation. Each pixel contains a logarithmic photoreceptor followed by a temporal differencing circuit. When the log-intensity change at pixel $(x, y)$ exceeds a contrast threshold $\pm C$, the pixel emits an event $e = (x, y, t, p)$ encoding position, microsecond-resolution timestamp, and polarity (brightness increase or decrease)~\cite{lichtsteiner2008dvs}. Pixels that see no change produce no output, yielding data rates proportional to scene dynamics rather than pixel count. Figure~\ref{fig:frame_vs_event} contrasts the output of a frame camera and a DVS observing the same rotating-disk scene; Figure~\ref{fig:bio_vs_neuromorphic} illustrates the biological-retina to memristive-synapse mapping that motivates the integration.

\subsubsection{Key specifications.}
This event-driven architecture confers three advantages over frame-based sensors. First, a dynamic range of 120--143~dB---roughly 60~dB wider than high-end CMOS imagers---enables operation from starlight to direct sunlight without exposure control (120~dB in the original DVS128~\cite{lichtsteiner2008dvs}; 143~dB in the ATIS QVGA~\cite{posch2011atis}). Second, sub-millisecond latency per event eliminates motion blur at high speeds. Third, sparse output reduces downstream bandwidth: a static scene generates zero data, whereas a 1280$\times$720 frame camera at 30~fps produces $\sim$27~MB/s regardless of content. These properties make DVS sensors attractive for latency-critical tasks such as obstacle avoidance, high-speed tracking, and always-on surveillance. The main limitations are the absence of absolute intensity information in pure DVS sensors, lower spatial resolution than mainstream frame cameras, and sensitivity to transistor mismatch that creates background activity noise.

\subsubsection{Sensor variants and state of the art.}
Three sensor families dominate the field: the original \emph{DVS} outputs events only~\cite{lichtsteiner2008dvs}; the \emph{DAVIS} adds a synchronous APS frame readout for co-registered intensity~\cite{brandli2014davis}; the \emph{ATIS} encodes absolute intensity as a time interval per event~\cite{posch2011atis}. Table~\ref{tab:dvs_sensors} summarizes the commercial and research sensors that anchor the maturity assessment in this review: Prophesee's Gen4 (IMX636, Sony fab) at $1280{\times}720$ with peak $\sim$1.066~Gev/s on a 4.86~\si{\micro\meter} pitch~\cite{finateu2020gen4}; the GenX320 at $320{\times}320$ in a $\sim$3~mW envelope (36~\si{\micro\watt} ultra-low-power)~\cite{prophesee2024genx320}; Samsung's Gen3 at $1280{\times}960$ with sub-5~\si{\micro\meter} pitch and 122~nW per-pixel power~\cite{suh2020samsung} (preceded by the $640{\times}480$ 300~Meps mass-production reference~\cite{son2017samsung}). Despite this progress, commercial DVS resolutions remain 4--16$\times$ lower than mainstream CMOS imagers and per-unit costs remain substantially higher, constraining adoption outside research and niche industrial settings.

\begin{table*}[t]
\centering
\caption{Specifications of Representative Event-Based Vision Sensors with Memristor Integration Prospects.}
\label{tab:dvs_sensors}
\resizebox{\textwidth}{!}{%
\begin{threeparttable}
\footnotesize
\setlength{\tabcolsep}{3pt}
\begin{tabular}{@{}l l c c c c c l@{}}
\toprule
\textbf{Sensor} & \textbf{Manufacturer} & \textbf{Resolution} & \textbf{Pixel Pitch} & \textbf{Dynamic Range} & \textbf{Latency} & \textbf{Power} & \textbf{Memristor Prospect} \\
\midrule
DVS128 \cite{lichtsteiner2008dvs}
  & iniVation & 128$\times$128 & 40~$\mu$m & 120~dB & $\sim$15~$\mu$s & $\sim$23~mW & R2 feasible (large pixel) \\

DAVIS \cite{brandli2014davis}
  & iniVation & 240$\times$180 & 18.5~$\mu$m & 130~dB & $<$1~ms\tnote{a} & $\sim$50~mW & R2 marginal \\

ATIS \cite{posch2011atis}
  & Austrian Inst. Tech. & 304$\times$240 & 30~$\mu$m & 143~dB & $<$3~$\mu$s & $\mu$W--mW\tnote{i} & R2 feasible (large pixel) \\

Gen3.1 VGA
  & Prophesee & 640$\times$480 & 15~$\mu$m & 120~dB & $<$100~$\mu$s & $\sim$300~mW typ.\tnote{f} & R2 marginal \\

Gen4.1 (IMX636) \cite{finateu2020gen4}
  & Prophesee/Sony & 1280$\times$720 & 4.86~$\mu$m & 124~dB\tnote{b} & $<$100~$\mu$s\tnote{c} & 5--205~mW\tnote{d} & R2 infeasible; R3 only \\

GenX320 \cite{prophesee2024genx320}
  & Prophesee & 320$\times$320 & 6.3~$\mu$m & $>$140~dB & $<$150~$\mu$s typ.\tnote{g} & 3~mW typ.\tnote{e} & R3 only \\

Samsung Gen2 \cite{son2017samsung}
  & Samsung & 640$\times$480 & 9~$\mu$m & 90~dB & \nr & \nr & R2 marginal \\

Samsung Gen3 \cite{suh2020samsung}
  & Samsung & 1280$\times$960 & 4.95~$\mu$m & 90~dB\tnote{h} & \nr & $\sim$27~mW typ. & R2 infeasible; R3 only \\
\bottomrule
\end{tabular}
\begin{tablenotes}
\small
\item R2 = in-pixel memristor memory; R3 = external crossbar accelerator.
\item[a] iniVation datasheet: $\sim$1~\si{\micro\second} temporal resolution; $<$1~ms worst-case latency.
\item[b] 124~dB corresponds to the HDR mode reported in Finateu ISSCC 2020; Prophesee/Sony product brief states $>$86~dB at 5~lux--100~klux, $>$120~dB at 80~mlux--100~klux.
\item[c] Typical latency $<$100~\si{\micro\second}; sub-10~\si{\micro\second} achievable only under high illumination.
\item[d] 5~mW standby to 205~mW peak operation per Prophesee product brief.
\item[e] GenX320 ultra-low-power mode: 36~\si{\micro\watt} per Prophesee product brief.
\item[f] Prophesee product documentation reports a 26--950~mW range across operating modes; $\sim$300~mW is the typical value.
\item[g] Prophesee GenX320 product brief: $<$150~\si{\micro\second} typical event latency; $\leq$1~ms minimum-contrast latency.
\item[h] Samsung Gen3 (Suh et al., ISSCC 2020): 90~dB dynamic range; fabricated in 28~nm FD-SOI BSI stacked CMOS.
\item[i] ATIS power is strongly activity-dependent: Posch~et~al.\ (JSSC~2011) report per-pixel and per-event energy figures rather than a single system-level wattage. The envelope spans tens of \si{\micro\watt} in low-activity regimes to a few mW under sustained event traffic, depending on array size and whether the grayscale-capture subsystem is active.
\item Modern sub-5~$\mu$m pixels (Gen4.1, Samsung Gen3, GenX320) cannot physically accommodate pixel-level 1D1M integration, forcing array-level (R3) or 3D-stacked approaches.
\end{tablenotes}
\end{threeparttable}%
}
\end{table*}

\subsection{Why Memristors for DVS?}
\label{sec:why-memristors}

Event cameras produce sparse, asynchronous, high-temporal-resolution data streams. Processing these streams on conventional von Neumann hardware forces serialization and dense memory access patterns that negate the sensor's efficiency. Several hardware technologies have been proposed to close this gap; Table~\ref{tab:competing_tech} provides a quantitative comparison along seven axes that govern DVS suitability: switching speed, endurance, retention, density, analog state support, CMOS compatibility, and an aggregated DVS-suitability rating.

Read across the rows, the table sorts the field into three regimes. The first regime---non-volatile analog crossbar candidates (RRAM/memristor, PCM, FeRAM, STT-MRAM)---offers fast switching ($<$100~ns) and multi-year retention with high BEOL-compatible density, but each has a single limiting factor: PCM drifts after programming and crystallisation currents are high, FeRAM is FEOL-bound and low-density, STT-MRAM is fundamentally binary, and RRAM/memristor variability is the highest of the four. The second regime---volatile / native-CMOS digital memory (SRAM, Flash, FPGA logic)---supports unlimited or very high cycle counts but cannot store synaptic weights without continuous power (SRAM, FPGA) or suffers prohibitive write times and limited endurance (Flash). The third regime---digital neuromorphic processors (Loihi~2, TrueNorth)---ships at milliwatt power and is the maturity benchmark, but stores weights in SRAM and performs no in-memory analog MVM. Memristor/RRAM is the only entry that combines analog multi-level states, non-volatility, high density, and BEOL compatibility; the rest of this section unpacks where this combination breaks down. The trade-offs that follow elaborate the row-by-row reasoning.

\subsubsection{Competing technologies and the memristor case.}
PCM offers fast switching ($\sim$50--100~ns) and has reached industrial-scale CIM deployment~\cite{ambrogio2023analogai, legallo2023mixedsignal}, but high programming currents and $\sim$$10^{8}$--$10^{9}$ endurance raise lifetime concerns for continuous event processing~\cite{sebastian2020memory}. CBRAM offers ultra-low operating currents but poor C2C repeatability and $\sim$$10^{6}$--$10^{8}$ endurance~\cite{ielmini2018memory}. STT-MRAM provides $>$$10^{12}$ endurance and CMOS compatibility but is fundamentally binary; FeRAM achieves low switching energy but faces scaling challenges below 28~nm and limited density~\cite{sebastian2020memory}. FPGAs offer mature dataflow at microsecond latency but consume 1--10~W with no in-memory analog MVM~\cite{kryjak2024event, karamimanesh2025snn_fpga}. Digital neuromorphic processors~\cite{davies2018loihi, merolla2014truenorth, furber2014spinnaker, richter2023speck} provide spiking support at milliwatt power but rely on SRAM-based synaptic storage. RRAM/memristors uniquely combine analog multi-level states, non-volatility, $<$10-nm footprint, and BEOL stackability above CMOS~\cite{ielmini2018memory}; projected energy is 1--10~fJ per MAC, two-to-three orders of magnitude below digital accelerators (\projected{}, not measured on a fabricated memristor-DVS system)~\cite{sebastian2020memory}. The cost: device variability degrades weight precision, IR drop limits practical arrays to $\sim$256--512 rows~\cite{hu2016dotproduct, chen2018neurosim}, sneak-path currents require complementary switches in passive arrays~\cite{linn2010complementary}, ADC/DAC peripherals dominate area and energy, and no fabricated system has integrated a memristor crossbar with a DVS sensor on a single die (though wafer-scale passive crossbar fabrication is demonstrated~\cite{choi2025waferscale}).

\subsection{The Case Against Memristors}
\label{sec:case-against}

A fair assessment must ask whether memristors are necessary at all.
SRAM-based compute-in-memory (CIM) at advanced nodes already achieves competitive energy efficiency: a mixed-precision SRAM-CIM and memristor-CIM processor demonstrated 40.91~TFLOPS/W on ResNet-20 (CIFAR-100) with $<$0.45\% accuracy degradation~\cite{khwa2025mixed_cim}, and a 28-nm bootstrapped-SRAM CIM macro reported 135.19~TOPS/W with layer-wise precision and sparsity~\cite{mao2024bootstrapped_sram}, matching or exceeding standalone memristor accelerators without the variability and endurance penalties.
Digital neuromorphic processors such as Intel Loihi~2 process DVS event streams at milliwatt power levels without any analog device non-idealities~\cite{davies2018loihi, orchard2021loihi2}.
The strongest counter-argument to memristors is maturity: SRAM-CIM and digital neuromorphic chips are shipping products, while the most advanced memristor-DVS prototype operates at 1,024-cell scale~\cite{huang2025multimode}.
Memristors retain two potential advantages---density (passive crossbars at 4F$^2$ vs.\ $>$100F$^2$ for SRAM) and non-volatile weight storage (eliminating reload from external memory)---but these advantages narrow as CMOS scaling continues and low-power SRAM retention improves. Replacing the silicon channel with graphene, which supports quasi-ballistic transport and higher mobility, can extend Moore's law incrementally without departing from the current integrated-circuit design paradigm~\cite{banadaki2016edge, banadaki2019graphene, banadaki2016thesis, banadaki2013novel}.
Whether the analog compute advantages of memristors justify their device-level challenges is an open question that the field has not yet answered experimentally for DVS applications. This question is not a peripheral caveat: it is the question that any honest assessment of memristor-DVS integration must answer first, and it shapes the evidence-graded analysis of Section~\ref{sec:applications} and the falsifiable roadmap of Section~\ref{sec:roadmap}.

\begin{figure*}[!htbp]
\centering
\includegraphics[width=0.85\textwidth]{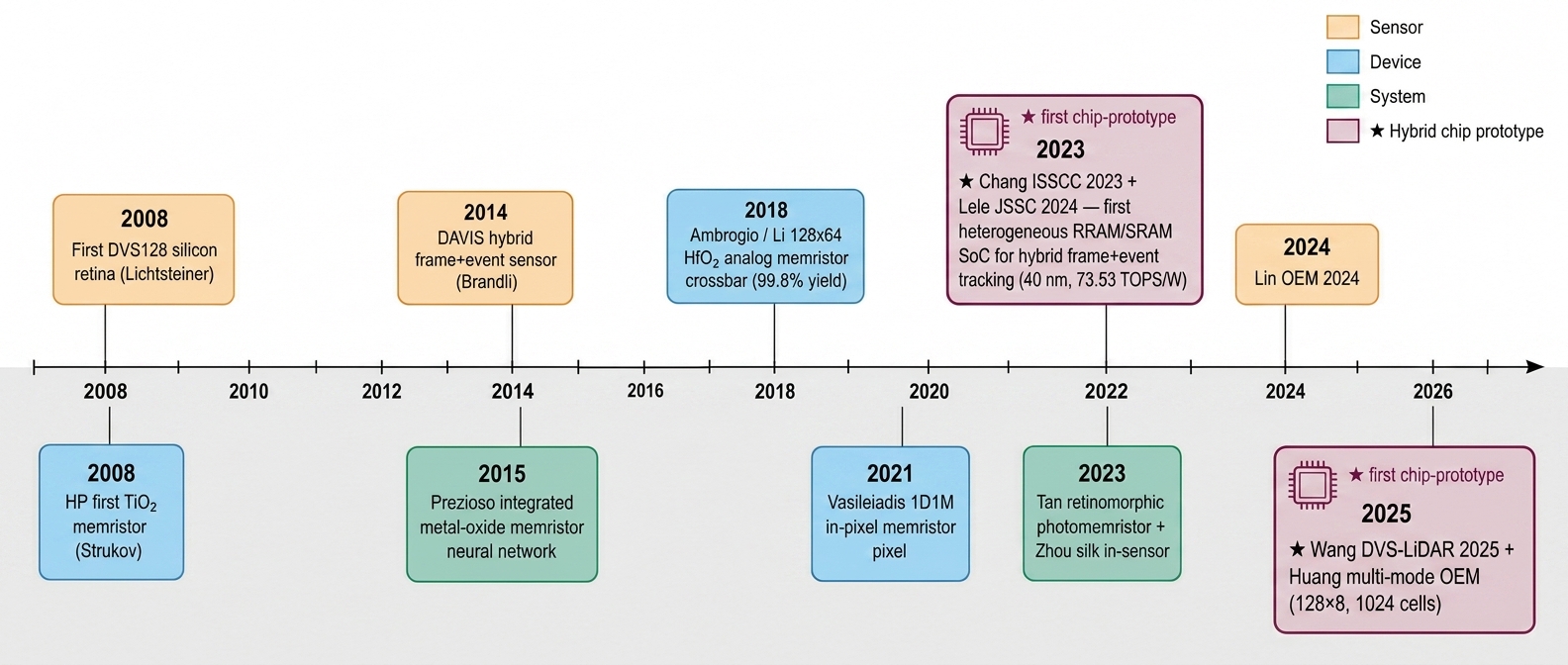}
\Description{Horizontal timeline 2008-2025 with nine milestones colour-coded by category (sensor, device, system). From left to right: 2008 first DVS128 silicon retina (Lichtsteiner), 2008 HP first TiO2 memristor (Strukov), 2014 DAVIS hybrid frame+event sensor, 2017 first memristor crossbar inference demo, 2018 Li et al. 128x64 HfO2 memristor crossbar at 99.8 percent yield, 2020 Prophesee Gen4 / Sony IMX636 1Mpx HD event camera, 2022 mature HfO2 / TaOx CMOS-compatible memristors, 2024 Yoon et al. simulated CMOS+memristor DVS hybrid, 2025 Huang et al. 128x8 fabricated multi-mode optoelectronic memristor array.}
\caption{Timeline of key milestones in memristor-based vision systems (2008--2025), colour-coded by category. The progression spans fundamental device physics~\cite{strukov2008missing} and the first DVS silicon retina, through the first memristor-crossbar inference demonstrations and the $128{\times}64$ HfO\textsubscript{2} crossbar at 99.8\% yield~\cite{li2018analogue}, megapixel commercial event cameras (Prophesee Gen4 / Sony IMX636), and recent system-level memristor-DVS efforts including the simulation-only CMOS--memristor hybrid~\cite{yoon2024memristor} and the fabricated $128{\times}8$ multi-mode optoelectronic memristor array of Huang et al.~\cite{huang2025multimode}.}
\label{fig:historical_timeline}
\end{figure*}

\paragraph{From foundations to the integration question.}
The device physics, sensor specifications, and competing-technology comparison above set the ground rules. A memristor-DVS proposal that ignores these constraints --- by, for example, claiming sub-microsecond DVS-rate operation on a perovskite stack with limited cyclic endurance, or projecting megapixel arrays from a 128$\times$8 demonstration --- is not credible at the device level. Figure~\ref{fig:historical_timeline} traces the parallel trajectories of memristor and DVS development that bring the two communities to the present integration question; Section~\ref{sec:architectures} then organises the fabricated literature into three integration roles (R1 photomemristor, R2 in-pixel, R3 crossbar accelerator) and analyses demonstrated performance against these ground rules at each level.

\begin{table*}[t]
\centering
\caption{Comparison of Memory and Processing Technologies for DVS Integration. Values represent typical reported ranges for mature device implementations.}
\label{tab:competing_tech}
\resizebox{\textwidth}{!}{%
\begin{threeparttable}
\footnotesize
\setlength{\tabcolsep}{3pt}
\begin{tabular}{@{}l c c c c c c c@{}}
\toprule
\textbf{Technology} & \textbf{Switch Speed} & \textbf{Endurance} & \textbf{Retention} & \textbf{Density} & \textbf{Analog States} & \textbf{CMOS Compat.} & \textbf{DVS Suitability} \\
\midrule
RRAM/Memristor
  & $<$100~ns & 10$^6$--10$^{12}$ & Years & High (4F$^2$) & Yes (multi-bit) & BEOL & High \\

PCM (Phase Change)
  & $<$50~ns & 10$^8$--10$^{12}$ & Years & High (4F$^2$) & Limited (drift) & BEOL & Moderate \\

STT-MRAM
  & $<$10~ns & $>$10$^{15}$ & Years & Moderate (30--60F$^2$) & No (binary) & BEOL & Low \\

FeRAM
  & $<$100~ns & 10$^{10}$--10$^{14}$ & Years & Low (22F$^2$) & Limited & FEOL & Low \\

SRAM (6T)
  & $<$1~ns & Unlimited & Volatile & Very low (100+F$^2$) & No & Native & Moderate \\

Flash (NAND)
  & $\sim\mu$s & 10$^3$--10$^5$ & Years & Very high & Yes (MLC) & Native & Low \\

FPGA
  & $<$10~ns & N/A & Volatile & Low & Digital only & N/A & Moderate \\

Loihi 2 (Intel)
  & $\sim\mu$s & N/A & Volatile & Moderate & Digital spikes & N/A & Moderate \\

TrueNorth (IBM)
  & $\sim$ms & N/A & Volatile & Low & Digital & N/A & Low \\
\bottomrule
\end{tabular}
\begin{tablenotes}
\small
\item F$^2$ = minimum feature size squared; BEOL = Back-End-Of-Line compatible; FEOL = Front-End-Of-Line.
\item DVS Suitability assessed based on: analog tunability for synaptic weights, non-volatility for background storage, density for pixel-level integration, and switching speed for event-rate processing.
\item RRAM/Memristor uniquely combines analog multi-level states, non-volatility, high density, and BEOL compatibility, though at the cost of higher variability and limited endurance compared to STT-MRAM.
\end{tablenotes}
\end{threeparttable}%
}
\end{table*}

\section{Architectural Paradigms and Integration}
\label{sec:architectures}

\begin{figure*}[!htbp]
\centering
\includegraphics[width=0.85\textwidth]{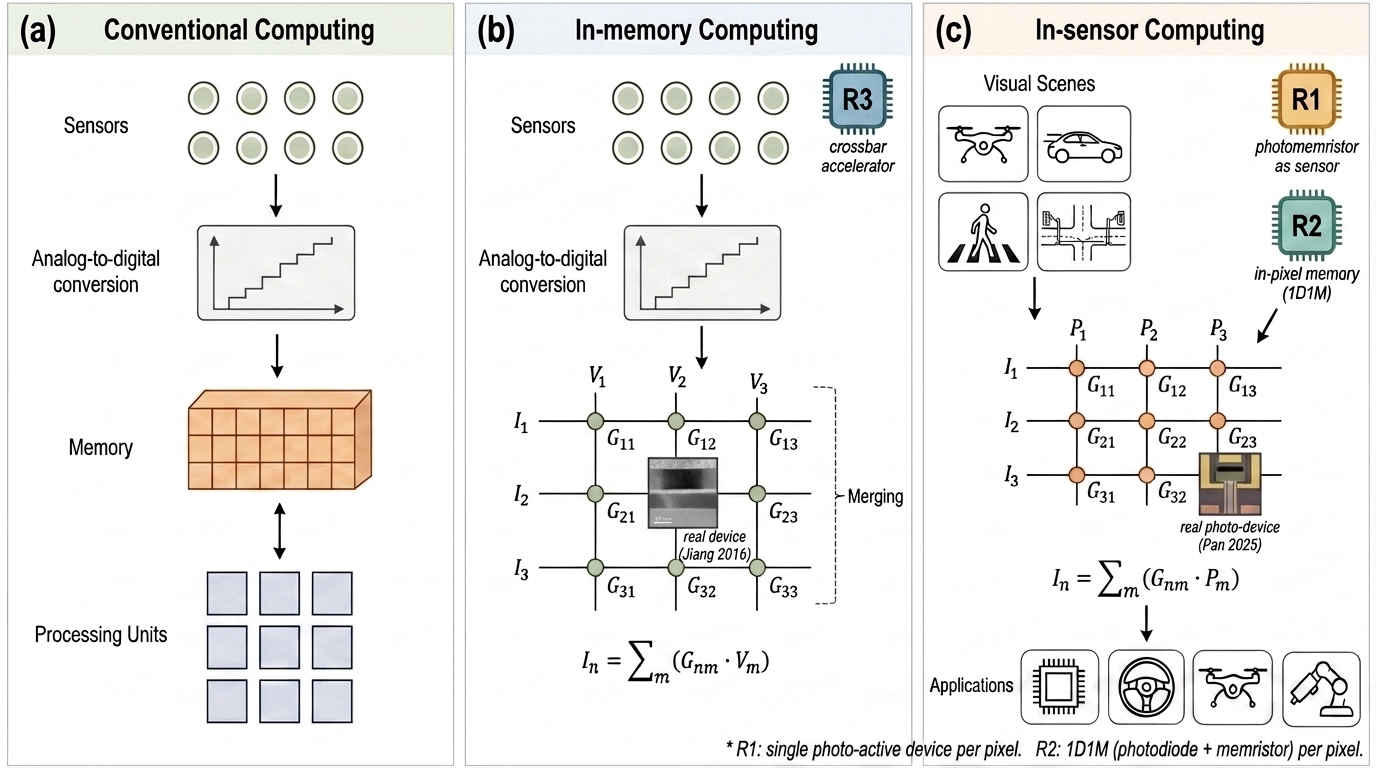}
\Description{Three-panel schematic of computing paradigms for DVS-relevant vision. Panel (a): conventional pipeline with separate sensor, ADC, memory, and processor blocks. Panel (b): in-memory computing pipeline that merges memory and processing into a memristor crossbar (Role 3). Panel (c): in-sensor computing where sensing, memory, and processing collapse into a memristor array fed directly by visual scenes (drone, car, pedestrian, traffic intersection) and physical stimuli, producing analog matrix-vector multiplication outputs that feed downstream applications (edge hardware, automotive ADAS, drone autopilot, surveillance / robotics).}
\caption{Three computing paradigms for vision, mapping onto the R1/R2/R3 taxonomy. (a)~Conventional pipeline with sensors, analog-to-digital conversion, memory, and processing units as separate blocks. (b)~In-memory computing merges memory and processing via a memristor crossbar (R3), while sensing remains separate. (c)~In-sensor computing merges all three stages: a reservoir-style memristor grid is driven directly by DVS-relevant visual scenes (drones, cars, pedestrians, traffic intersections) and physical stimuli (temperature, pressure, motion, magnetic field), with outputs feeding edge-deployment applications (R1, photomemristor; R2, in-pixel memory). Adapted from Chen et al., \emph{npj Unconv.\ Comput.} \textbf{2}, 19 (2025)~\cite{chen2025twodim} under CC BY 4.0; redrawn with DVS-relevant input scenes in panel (c).}
\label{fig:integration_paradigms}
\end{figure*}

\begin{figure*}[!htbp]
\centering
\includegraphics[width=0.7\textwidth]{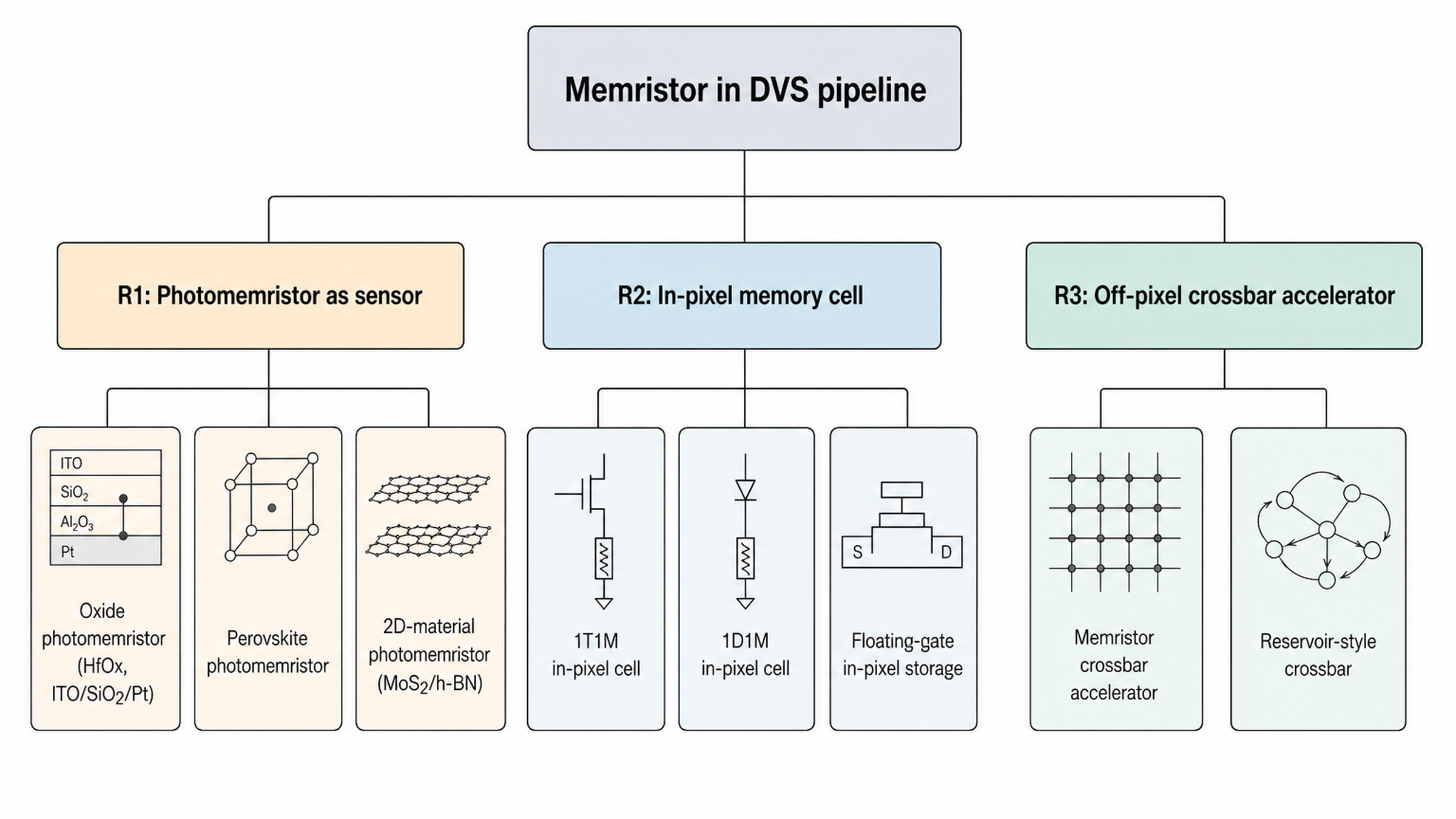}
\Description{Hierarchical three-branch taxonomy tree. Root: Memristor in DVS pipeline. Three Level-1 branches (R1, R2, R3) fan downward to sub-categories with realistic mini-icons: R1 Photomemristor as sensor (oxide HfOx ITO/SiO2/Pt, perovskite, 2D-material MoS2/h-BN); R2 In-pixel memory cell (1T1M, 1D1M, floating-gate storage); R3 Off-pixel crossbar accelerator (memristor crossbar, reservoir-style crossbar).}
\caption{Taxonomy of memristor integration roles in DVS systems, refining Figure~\ref{fig:integration_paradigms} with device-specific examples. Three architectural paradigms (R1/R2/R3) are identified, each with different device requirements, maturity levels, and trade-offs; sub-branches list the representative material families and device topologies surveyed in this section.}
\label{fig:taxonomy}
\end{figure*}

The development of memristor-based vision systems unfolds across three eras: foundational device physics (2008--2014), array-scale demonstrations (2015--2020), and the current era of system-level integration (2021--2025), as illustrated in Figure~\ref{fig:historical_timeline} (Section~\ref{sec:foundations}).

Before detailing the integration roles, Figure~\ref{fig:integration_paradigms} situates them within the broader shift from conventional von-Neumann pipelines to in-memory and in-sensor computing.
Memristors intersect with event-based vision sensors in three distinct roles, summarized in Figure~\ref{fig:taxonomy}.
First, a \emph{photomemristor} replaces the photodiode entirely, combining
light sensing and analog memory in a single device~\cite{cao2023dual,cai2023insitu,tan2023dynamic};
recent extensions of this paradigm have demonstrated multispectral
color-discriminating optoelectronic synapses suitable for ultra-high-resolution
vision chips~\cite{lee2023multispectral} and hemispherical retinomorphic arrays
that realise wide field-of-view stereo vision through binocular
disparity~\cite{shan2024hemispherical}; complementary device-level reviews
catalogue the broader optoelectronic-memristor landscape across metal-oxide,
low-dimensional, polymer, and bio-inspired stacks~\cite{shrivastava2025neuromorphic, yang2026metaloxide, pei2026lowdim_opto, zhou2025polymer_memristor, wang2025bioinspired_opto}, and bio-inspired in-materia analog photoelectronic reservoirs extend R1-style devices to richer temporal computation~\cite{cui2025photoelectronic, li2024neuromorphic_reservoir}.
Second, an \emph{in-pixel memristor} sits alongside a conventional photodiode
in a one-diode--one-memristor (1D1M) cell, storing per-pixel state to enable
local change detection~\cite{vourkas2021inmemory,laiho2015memristive}; an
explicitly retinomorphic instantiation has been demonstrated using 1T1M
photoreceptor and ganglion cells that perform shape extraction in
sensor~\cite{bao2018shape}, and bioinspired retinal architectures emulating
horizontal and amacrine cells through memristive layers have been integrated in
180-nm CMOS~\cite{eshraghian2018nvh}.
Third, a \emph{crossbar accelerator} processes the event stream after it
leaves the sensor, performing matrix-vector multiply (MVM) in the analog
domain~\cite{huang2025multimode,li2018analogue,yoon2024memristor}; early
end-to-end demonstrations included a CMOS image sensor driving a PCMO memristor
array with on-chip neurons for digit recognition~\cite{chu2015neuromorphic}, and
more recent retinomorphic memristive crossbars have shown analog-regime letter
recognition and object tracking~\cite{wang2020retinomorphic}.
Each role imposes different constraints on switching speed, endurance,
variability, and CMOS compatibility.
This section examines all three roles through the lens of ten core
references that define the current state of memristor--DVS integration.
Four cross-cutting patterns frame the discussion (the endurance--photosensitivity trade-off, the single-device-to-array-scale gap at the DVS intersection, the absence of end-to-end system-level metrics, and reporting incompleteness across the corpus) and are quantified with citations in the comparative analysis (Section~\ref{sec:comparative}).

\subsection{The Interface Challenge: Event-to-Memristor Encoding}
\label{sec:interface}

Dynamic vision sensors produce asynchronous address-event representation streams (pixel address, microsecond timestamp, polarity)~\cite{lichtsteiner2008dvs}, while memristive crossbar arrays require analog input voltages applied simultaneously to all rows for a single MVM step~\cite{li2018analogue}. Bridging these two data formats is the first engineering challenge in any memristor--DVS system, and three encoding strategies have emerged in the literature.

\emph{Rate coding} bins events into fixed-duration windows and maps the event count per pixel to a proportional input voltage; Huang et al.\ drive a 128$\times$8 OEM array this way and achieve 91.2--96.1\% accuracy on motion and pattern classification~\cite{huang2025multimode}, but a 60\,Hz binning rate (as in~\cite{tan2023dynamic}) collapses the microsecond temporal resolution of the DVS to the millisecond regime, eliminating the latency advantage that motivates event cameras. \emph{Temporal coding} preserves per-event timing by mapping the inter-spike interval to pulse width or amplitude; ferroelectric organic devices~\cite{cai2023insitu} and the $>$$10^9$-cycle perovskite ferroelectric of~\cite{wang2024perovskite} are durability-compatible with the sustained write rates DVS-driven inference would impose, but no published work has demonstrated closed-loop temporal coding from a physical DVS output to a memristor input at full event bandwidth. \emph{Spike-based direct coupling} avoids analog conversion entirely by treating each DVS event as a binary spike modulating memristor conductance; the oxide pixel of~\cite{laiho2015memristive} performs background subtraction at sub-millisecond update rates this way and is the most compatible with SNN back-ends, but requires write-cycle endurance the photomemristor-class pixel devices have not demonstrated under sustained continuous operation~\cite{vourkas2021inmemory, cao2023dual}.

\subsection{Pixel-Level Integration}
\label{sec:pixel-level}

Pixel-level integration places the memristor inside the pixel circuit,
enabling per-pixel analog memory without off-chip data transfer.
Two variants exist: the photomemristor, which combines sensing and memory
in one device, and the 1D1M cell, which pairs a conventional photodiode
with a co-located memristive element.

\subsubsection{The 1D1M Architecture}

The SiN\textsubscript{x}-based 1D1M pixel concept of Vasileiadis et al.~\cite{vourkas2021inmemory}
demonstrates (in SPICE simulation, based on experimentally extracted device
parameters) 8 distinguishable resistance levels at the photodiode--memristor pixel.
The device structure uses a CMOS-compatible back-end-of-line (BEOL) fabrication
process, making it integrable with existing foundry flows.
Two limitations constrain its applicability to event-driven vision.
First, as an architectural study the paper does not report experimental
endurance; sustained event-driven updates typically demand
$10^6$--$10^9$ cycles.
Second, the authors do not quantify fill-factor impact; adding a memristor
stack to the pixel reduces the photosensitive area, potentially degrading
sensitivity in low-light conditions where event cameras already struggle.
Zhou et al.~\cite{zhou2023insensor} demonstrate a complementary architecture: an event-driven sensor feeding an in-sensor SNN, but using photodiodes rather than memristors. The R1/R2/R3 taxonomy is defined by where the \emph{memristor} sits; this design lies outside it. It nonetheless sets the digital baseline that any future Role~2/R3 prototype must beat.

The oxide-memristor pixel of~\cite{laiho2015memristive} targets a different
function: per-pixel background subtraction.
By storing a reference conductance state, the device flags deviations that
exceed a threshold, functionally equivalent to the change-detection
comparator in a DVS pixel.
The sub-millisecond update time and low per-pixel power consumption
are promising, but neither endurance data nor array-scale uniformity
measurements have been reported in the accessible version of the paper.
More recently, Lin et al.~\cite{lin2025retinomorphic} fabricated an event-driven retinomorphic photodiode with bio-plausible temporal dynamics in a vertically integrated organic heterojunction, producing spiking outputs on brightness change, functionally analogous to a DVS pixel in a single Role~1 device.

\subsubsection{Photomemristors as Event Sensors}

Photomemristors eliminate the photodiode entirely by using light-modulated
resistive switching to perform both sensing and memory storage.
The perovskite dual-functional device of~\cite{cao2023dual}
achieves 276\,mA/W responsivity and $4.7 \times 10^{11}$\,Jones
detectivity, competitive with commercial InGaAs detectors, while operating
at only $3 \times 10^{-11}$\,W with a 0.15\,V setting voltage and
$>$7000\,s retention.
These operating-power figures are orders of magnitude below typical
CMOS comparator dissipation, suggesting the potential for radical power savings
at the pixel level if cycling endurance can be addressed.

Two caveats apply: ``good cyclic stability'' has not been characterised at the cycle counts repetitive event-driven operation requires, and perovskite stability under humidity and thermal cycling (a known concern in the photovoltaics literature~\cite{niu2015perovskite}) remains uncharacterised for this device.

The ferroelectric organic device of~\cite{cai2023insitu} takes a different approach: polarisation-state accumulation over hours enables motion history encoding without external buffers, but switching speed, write endurance, and UV stability remain uncharacterised~\cite{jorgensen2008stability}.

\paragraph{Ferroelectric HfO\textsubscript{2} and CMOS-compatible multilevel weights.}
Doped ferroelectric HfO\textsubscript{2} (HfZrO, Si-doped) supports polarisation-based multilevel weights at advanced CMOS nodes without additional photolithography masks~\cite{martemucci2025ferroelectric}; perovskite ferroelectric photomemristors~\cite{wang2024perovskite} extend this logic into the R1 role; the optoelectronic-RRAM of Zhou et al.~\cite{zhou2019optoelectronic} is the closest oxide predecessor of a neuromorphic R1 element. None has been coupled to a DVS sensor in a fabricated system.

Figure~\ref{fig:memristor_switching}(d--f) summarises the device-physics picture for a representative oxide photomemristor: the ITO/SiO\textsubscript{2}/Al\textsubscript{2}O\textsubscript{3}/Pt stack with its 405\,nm illumination protocol and ion/vacancy distribution (d), the light-modulated I--V response under illumination (e), and the atomistic oxygen-vacancy mechanism that connects the photon flux to the conductance change (f).

\subsubsection{The Fill-Factor--Complexity Trade-off}

Any in-pixel device competes for area with the photosensitive element: a 1D1M cell adds at minimum a select transistor and a memristor stack to a DVS pixel already containing a logarithmic photoreceptor, differencing amplifier, and comparator. No published 1D1M design has reported fill-factor measurements, making the sensitivity penalty impossible to assess. Recent in-pixel demonstrations extend this direction: one-pixel-multiple-memristor architectures~\cite{sun2026one_pixel}, memristive cellular neural networks~\cite{ravichandran2026cellular}, and neuro-memristive HDR sensors~\cite{paissan2024hdr}.

\subsection{Array-Level Processing and Crossbar Accelerators}
\label{sec:array-level}

In the crossbar accelerator role, memristors sit after the sensor and perform in-memory matrix-vector multiplication on the event stream, decoupling the sensing problem from the computing problem and allowing mature DVS sensors to pair with separately optimized memristive arrays.

\subsubsection{Memristor-CIM at Array Scale: Current SOTA and the DVS Gap}

Broader memristor compute-in-memory (CIM) literature has advanced substantially beyond the small-array results dominating the memristor--DVS intersection: TRL~6--7 demonstrations now span Rao et al.'s $256{\times}256$ HfO\textsubscript{2} array with 2{,}048 analog conductance levels~\cite{rao2023thousands}, NeuRRAM's 48-core 3-million-cell RRAM-CIM chip~\cite{wan2022neurram}, IBM HERMES~\cite{ambrogio2023analogai}, and Le Gallo et al.'s 64-core mixed-signal CIM chip~\cite{legallo2023mixedsignal}. With the partial exception of the Chang/Lele 2023/2024 SoC~\cite{sharma2023isscc, sharma2024jssc} (which fuses a 40-nm RRAM CIM macro with an SRAM-based SNN co-processor for hybrid target tracking), none has been coupled to a DVS sensor or benchmarked on an event-camera task, and even Chang/Lele routes the event pipeline through SRAM rather than memristor crossbars. The gap is therefore not a device-physics gap but an \emph{integration-and-interface} gap; SpiDR~\cite{sharma2024spidr} pursues the same target with a digital substrate, and Zhang et al.~\cite{zhang2026ssm_cim} have begun the analog path with a CIM implementation of state-space models for event-sequence processing. Any DVS-specific memristor-crossbar proposal is now held to two simultaneous benchmarks: the measured performance of these non-DVS RRAM-CIM chips on similar-complexity inference tasks, and the measured performance of digital neuromorphic processors (Loihi~2, SynSense Speck) that already process DVS event streams today.

Within the DVS-adjacent literature, the HfO\textsubscript{2}-based $128{\times}64$ 1T1R crossbar of~\cite{li2018analogue} achieves 99.8\% device yield and 5--8-bit equivalent analog precision per cell, sufficient for convolutional inference under post-training quantization~\cite{gholami2021survey}; on-chip memristor training has been demonstrated in a fabricated 30-device prototype at 97\% classification accuracy under realistic device variations~\cite{eslami2024onchip}. These DVS-adjacent demonstrations remain orders of magnitude smaller than Rao~2023 or NeuRRAM in cell count, have not been tested on event-camera data, and require DAC conversion of the event stream.

\subsubsection{Optoelectronic Memristor Arrays}

The multi-mode optoelectronic memristor (OEM) array of~\cite{huang2025multimode} addresses the conversion bottleneck by accepting optical inputs directly. Operating in both volatile and non-volatile modes, the 128$\times$8 array classifies motion patterns and static images with 91.2--96.1\% accuracy while consuming 20$\times$ less energy than a GPU baseline (uncontrolled for model complexity: single-layer OEM versus deeper GPU network).

Two further qualifications apply: the 1{,}024-cell array requires $10^3\times$ scaling to reach megapixel DVS densities~\cite{finateu2020gen4}, and volatile mode limits temporal context for event processing.

\subsubsection{Retinomorphic Reservoir Computing (Role~1 with reservoir readout)}

The retinomorphic sensor of~\cite{tan2023dynamic} is a Role~1 photomemristor operated as a reservoir computer rather than a classical per-pixel sensing element. A photomemristive element with ON/OFF ratio on the order of $10^2$ generates nonlinear dynamics that encode temporal patterns intrinsically; only the digital readout layer is trained. The device was tested at optical-input frequencies up to 60\,Hz (matched to commercial display rates), and on this input it performs motion recognition and short-term prediction. The assignment below (and in Table~\ref{tab:performance}) classifies this work as R1 for taxonomy purposes, with the reservoir-computing paradigm noted as its architectural specialisation rather than as a separate crossbar-accelerator role.

The 60\,Hz tested input is three to four orders of magnitude below DVS event rates, limiting applicability to scenes where temporal down-sampling is acceptable. Whether memristive reservoirs can match trained deep networks on complex event-camera tasks (multi-class detection, optical flow) remains open.

\subsection{Hybrid CMOS-Memristor Systems}
\label{sec:hybrid}

Hybrid architectures retain CMOS for control and digital logic while offloading analog computation to memristive arrays.

\subsubsection{Simulation-Based Hybrids}

Yoon et al.~\cite{yoon2024memristor} simulated a hybrid CMOS-memristor system on POKER-DVS and MNIST-DVS, reporting 79\% / 75\% power savings at 0.5\% / 0.75\% accuracy degradation relative to a pure-CMOS baseline (\simulated{}). The system has not been fabricated, and simulation power-savings estimates routinely diverge from silicon measurements due to parasitic, wire-resistance, and peripheral overheads simulations underestimate~\cite{xia2019memristive}. Even if the 75--79\% reduction is optimistic by a factor of two, the architectural template (CMOS control with memristive MAC units) merits fabrication-level validation.

\subsubsection{Bio-Inspired Dual-Mode Architectures}

The silk-protein device of~\cite{zhou2023silk} implements ORRAM (light-driven sensing) and ERRAM (computation) on the same substrate, mirroring the retina-to-cortex signal chain. Silk fibroin biocompatibility is a genuine differentiator for implantable applications, but endurance and scalability data are absent: whether silk memristors can achieve the $10^6$+ cycles and sub-percent variability required for practical vision processing remains unknown. Across all three hybrid variants, no published work has fabricated a complete pipeline from DVS event generation through memristive processing to task-level output on a single chip or board, illustrating the central barrier to validating the performance claims that motivate this research direction.

\subsection{Advanced Material Systems}
\label{sec:materials}

\begin{figure*}[!htbp]
\centering
\includegraphics[width=0.95\textwidth]{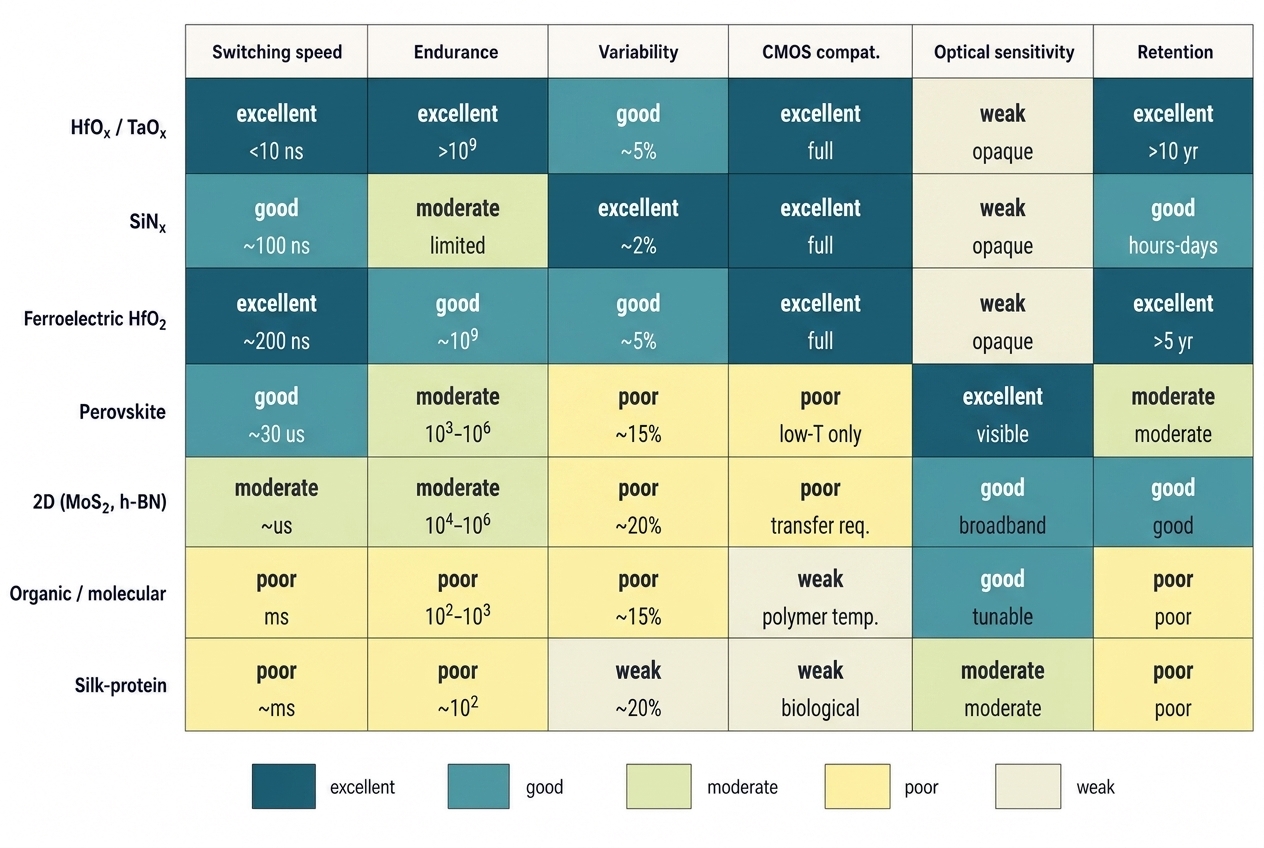}
\Description{A 7x6 colour-coded categorical heatmap. Rows: seven candidate material systems (HfOx/TaOx, SiNx, ferroelectric HfO2, perovskite, 2D MoS2/h-BN, organic/molecular, silk-protein). Columns: six DVS-relevant requirements (switching speed, endurance, variability, CMOS compat., optical sensitivity, retention). Each cell is filled with a soft tier colour and contains both a tier word ("excellent", "good", "moderate", "poor", "weak") and a representative value. A horizontal legend at the bottom maps the five tiers from soft sage (excellent) through pale sage, warm sand, muted apricot, to muted rose (weak). No row is uniformly excellent across all six requirements.}
\caption{Material-system compatibility matrix for memristor-based DVS integration. Rows are seven candidate material systems; columns are six DVS-relevant requirements. Cells are scored on a five-tier categorical scale (excellent / good / moderate / poor / weak). Each cell carries both the tier word and a representative value (e.g., ``$<$10\,ns'' for switching speed). No material simultaneously achieves the top tier across all six requirements; per-paper quantitative values appear in Supplement~S4.}
\label{fig:material_heatmap}
\end{figure*}

The choice of resistive switching material determines every device-level
metric: switching speed, endurance, variability, retention, CMOS
compatibility, and optical responsivity (for photomemristors).
Figure~\ref{fig:material_heatmap} summarises the compatibility landscape across the seven candidate material systems discussed below; per-paper quantitative values appear in Supplement~S4.

The prevalence of NR (not reported) cells in the per-paper supplement table mirrors Pattern~4 in Section~\ref{sec:comparative}: cross-architecture comparison rests on incomplete data, and the benchmark specification proposed in Section~\ref{sec:standardization} addresses this directly.

\subsubsection{Oxide-Based Materials}

Metal oxides (HfO\textsubscript{2}, TaO\textsubscript{x}, SiN\textsubscript{x}) offer the strongest path to CMOS integration because they are already present in standard foundry processes; HfO\textsubscript{2} is used as a high-$\kappa$ gate dielectric at the 28\,nm node and below, allowing 1T1R arrays to be fabricated without exotic deposition steps~\cite{li2018analogue}. SiN\textsubscript{x} is similarly BEOL-compatible, though its lower endurance limits it to applications with infrequent weight updates~\cite{vourkas2021inmemory}. The primary DVS-integration limitation is the lack of intrinsic photosensitivity: crossbar-role oxide devices require a separate sensor front-end and DAC conversion, adding area and power overhead.

\subsubsection{Perovskite-Based Materials}

Halide perovskites combine optical absorption with resistive switching, enabling true photomemristive operation. The perovskite ferroelectric device of~\cite{wang2024perovskite} combines $>$$10^9$ polarization-reversal cycles, multi-day retention, 4-bit analog precision, and high cycle-to-cycle linearity. These metrics, if reproducible at array scale, would satisfy core requirements for event-driven vision processing, but most reported metrics come from single-device characterization, and perovskite films exhibit grain-boundary-dependent switching variability that worsens with array size~\cite{niu2015perovskite}. The qualitative cyclic-stability characterization of the non-ferroelectric perovskite device~\cite{cao2023dual} suggests not all perovskite formulations achieve the endurance of the ferroelectric variant. Zhang et al.~\cite{zhang2026phototriggered} recently demonstrated a monolithic 2D-perovskite photomemristor array with pW-level in-sensor computing, $\sim$$10^5$ ON/OFF ratios, and $>$22{,}000~s retention, a concrete step toward a fully integrated Role~1 vision front-end. Environmental stability under humidity, thermal cycling, and prolonged illumination remains the primary concern for perovskite deployment outside laboratory conditions.

\subsubsection{Molecular-Ferroelectric and Biological Materials}

Molecular ferroelectric photomemristors~\cite{cai2023insitu} and silk-protein composites~\cite{zhou2023silk} occupy the far end of the CMOS-compatibility spectrum: neither is foundry-integrable, and both target niche applications where biocompatibility or mechanical flexibility outweigh fabrication compatibility. The molecular-ferroelectric device contributes hours-long decay constants that store motion history without external circuitry, though switching speed and endurance remain uncharacterised and organic/polymeric stacks degrade under UV and thermal stress~\cite{jorgensen2008stability}. Silk fibroin offers proven biocompatibility plus dual-mode (ORRAM/ERRAM) operation for implantable retinal-prosthesis directions, but quantitative benchmarking against inorganic alternatives remains open.

\subsection{Comparative Analysis}
\label{sec:comparative}

\begin{table*}[t]
\centering
\caption{Comparative Performance Metrics of Representative Memristor-DVS Implementations. \nr{} = Not Reported. All values are as reported in the original studies; direct cross-study comparison requires caution due to differing experimental conditions.}
\label{tab:performance}
\resizebox{\textwidth}{!}{%
\begin{threeparttable}
\scriptsize
\setlength{\tabcolsep}{3pt}
\begin{tabular}{@{}l l l l l l l l@{}}
\toprule
\textbf{Work} & \textbf{Role} & \textbf{Switching Speed} & \textbf{Endurance} & \textbf{Power} & \textbf{Retention} & \textbf{Variability} & \textbf{Key Metric} \\
\midrule
\cite{vourkas2021inmemory} SiNx 1D1M
  & R2 & $\sim$kHz update & \nr\tnote{a} & \nr & \nr & SPICE-simulated\tnote{b} & 8 levels (SPICE) \\

\cite{laiho2015memristive} Oxide pixel\tnote{c}
  & R2 & $<$1~ms update & \nr & Low (per-pixel) & \nr & \nr & Background subtraction \\

\cite{cao2023dual} Perovskite
  & R1 & \nr & \nr (good cyclic stability) & $3\times10^{-11}$~W & $>$7000~s & Stable & 276~mA/W; $4.7\times10^{11}$~Jones \\

\cite{cai2023insitu} Organic
  & R1 & \nr & \nr & 54.55~$\mu$W/cm$^2$ & Hours (decay) & \nr & Temporal integration \\

\cite{wang2024perovskite} Perovskite ferro.
  & R1 & \nr & $>$10$^9$ cycles & \nr & $>$10$^4$~s & \nr & Polarisation switching \\

\cite{huang2025multimode} Multi-mode OEM
  & R1 (w/ digital readout)\tnote{e} & \nr & \nr & 20$\times$ less vs.\ GPU & Non-volatile & \nr\tnote{d} & 91.2--96.1\% acc. \\

\cite{tan2023dynamic} Retinomorphic
  & R1 & 60~Hz frame rate & \nr & In-sensor & Per-frame & Uniform & ON/OFF $>$10$^2$ \\

\cite{li2018analogue} HfO$_2$ crossbar
  & R3 & $\sim\mu$s MAC & \nr & $\ll$ digital & Non-volatile & 99.8\% yield & 128$\times$64; 5--8~bit \\

\cite{yoon2024memristor} Hybrid CMOS\tnote{g}
  & R3 & Event-driven & \nr & 75--79\% saved\tnote{g} & \nr & 0.5--0.75\% acc.\ loss\tnote{g} & POKER-DVS; MNIST-DVS \\

\cite{zhou2023silk} Silk-protein
  & Dual-mode R1+R3\tnote{f} & Parallel & \nr & Low (edge) & Non-volatile & \nr & Dual-mode ORRAM+ERRAM \\
\bottomrule
\end{tabular}
\begin{tablenotes}
\small
\item Role: R1 = photomemristor (sensor), R2 = in-pixel memory, R3 = crossbar accelerator.
\item[a] Vasileiadis et al.\ 2021 is a SPICE-based architectural study; experimental endurance is not reported.
\item[b] The paper reports SPICE-simulated level-to-level variability for the 8 distinguishable resistance states extracted from a data-fitted behavioural model; specific numerical bounds are documented in the paper body.
\item[c] Laiho \& Lehtonen 2015 is a 4-page ISCAS description of the pixel topology; the memristor material is not explicitly stated in the accessible version.
\item[d] Huang et al.\ 2025 report high array-level yield for the 128$\times$8 OEM array; a specific percentage is reported in the supplementary information rather than the main text.
\item[e] The OEM array accepts optical input (Role~1 photomemristor) and performs in-sensor matrix operations, but its output is digitised through an external ADC/processor rather than routed to a second memristor crossbar --- it is therefore classified as Role~1 with a digital readout rather than a joint R1/R3 system.
\item[f] Zhou et al.\ 2023 silk-protein devices operate in two distinct regimes (ORRAM for preprocessing, ERRAM for classification) on the same substrate, functionally spanning both the photomemristor (R1) and crossbar-accelerator (R3) roles within a single material platform.
\item[g] Yoon et al.\ 2024 is a circuit-simulation study; the 75--79\% power saving and 0.5--0.75\% accuracy loss (POKER-DVS / MNIST-DVS respectively) are simulated (\simulated{}), not measured on fabricated hardware.
\item Of the quantitative entries expected in this table, approximately one third contain absolute values with units; the remainder are qualitative or unreported, reflecting the early maturity of this field.
\item \textbf{Takeaway:} of 80 expected metric cells (8 metrics $\times$ 10 works), 50 are \nr{}; cross-architecture comparison on common metrics is currently impossible. This reporting gap is itself a finding (see Pattern~4 in Section~\ref{sec:comparative}) and motivates the joint-metric specification in Section~\ref{sec:standardization}.
\item \textbf{Heterogeneous units.} Several columns aggregate values reported in non-comparable units, reflecting how each cited work chose to report. The Power column mixes absolute power ($3\!\times\!10^{-11}$~W), areal power density (54.55~$\mu$W/cm$^2$), GPU-relative comparisons (20$\times$ less), pure-CMOS-relative power-saving fractions (75--79\% saved), and per-pixel/per-edge qualitative descriptors. The Switching-Speed column similarly mixes update rates (kHz), settling times ($<$1~ms), frame rates (60~Hz), and event-driven (asynchronous) regimes. Cross-row comparison requires reading the metric type alongside the figure; the joint-metric specification proposed in Section~\ref{sec:standardization} (B5) is the prerequisite for collapsing these columns into directly comparable units.
\end{tablenotes}
\end{threeparttable}%
}
\end{table*}

Table~\ref{tab:performance} compiles the quantitative performance metrics
across all ten core references, organized by memristor role.
Four patterns emerge from the comparison.

\paragraph{Pattern 1: The endurance--photosensitivity trade-off.}
Oxide-based devices (HfO\textsubscript{2}, SiN\textsubscript{x}) achieve
CMOS compatibility and, in the case of HfO\textsubscript{2}, $>10^6$-cycle
endurance, but they cannot sense light.
Photomemristors (perovskite, organic) sense light directly but, with the
exception of the perovskite ferroelectric~\cite{wang2024perovskite}, exhibit limited cyclic endurance under continuous switching, well below the $>$$10^9$ levels routinely reported for oxide and ferroelectric stacks.
The field needs either high-endurance photomemristors or zero-overhead
optical-to-electrical conversion for oxide crossbars.

\paragraph{Pattern 2: Single-device versus array-scale validation.}
The best-performing devices are characterized at the single-device level.
The perovskite ferroelectric~\cite{wang2024perovskite} reports
sub-percent variability, but only between two terminals on one device.
The OEM array~\cite{huang2025multimode} scales to 1,024 cells and the
HfO\textsubscript{2} crossbar~\cite{li2018analogue} to 8,192
cells (128$\times$64); no other core reference exceeds single-device
or few-device demonstrations.
Megapixel-scale integration (the minimum for practical DVS applications) remains undemonstrated.

\paragraph{Pattern 3: Missing system-level metrics.}
None of the ten core references report end-to-end system metrics:
frames-per-second throughput, detection accuracy on standard benchmarks
(e.g., Gen1, 1Mpx, DSEC), or total system power including peripherals.
The reported metrics (responsivity, variability, endurance) are device-level figures of merit that do not directly predict system performance.

\paragraph{Pattern 4: Reporting incompleteness across the corpus.}
An energy--latency comparison across the ten core references would
provide the most direct assessment of architectural trade-offs.
However, of the ten works, only three report \emph{both} latency and energy
metrics with quantitative values; the remaining entries report one metric or neither, as reflected by the \nr{} entries in Table~\ref{tab:performance}.
Roughly 62\% of the expected joint-metric data points are missing, a finding that directly motivates the benchmark specification proposed in Section~\ref{sec:standardization}.

\subsection{The ADC/DAC Bottleneck}
\label{sec:adc-bottleneck}

Analog-to-digital and digital-to-analog converters (ADCs/DACs) mediate
every interface between digital event streams and analog memristive
arrays.
Their overhead threatens to negate the energy and area advantages of
in-memory computing.

\subsubsection{The Scale of the Problem}

In memristive accelerator prototypes, peripheral circuits (ADCs, DACs, sample-and-hold stages, sense amplifiers) dominate both area and energy. The ISAAC architecture analysis~\cite{shafiee2016isaac} reports peripheral circuitry consuming roughly half of tile energy and a third of tile area, with ADCs alone the single largest contributor; the Xia and Yang survey~\cite{xia2019memristive} confirms peripherals dominate area and energy in current accelerators. Le Gallo et al.~\cite{legallo2023mixedsignal} replace successive-approximation ADCs with charge-based time-encoded read-out, reducing this overhead at the cost of added latency on long accumulation columns; Rao et al.~\cite{rao2023thousands} achieve 2{,}048 conductance levels per device but do not publish a tile-level area/energy decomposition. The ISAAC ratio is therefore treated as the canonical \emph{order-of-magnitude} reference, sensitive to read-out architecture and precision. For the noise-limited regime, each additional bit of ADC resolution increases power $\sim$$4\times$~\cite{shafiee2016isaac, xia2019memristive}, creating an exponential penalty for high-precision analog computation. The precision mismatch compounds the problem: a 5-bit memristor (32 levels) in a 256-row crossbar produces a column dynamic range requiring $\log_2(32\times 256) = 13$ bits to resolve, far exceeding the device's native precision.

\subsubsection{Mitigation Strategies}

Three approaches reduce or eliminate the ADC bottleneck. \emph{Architectural innovation:} MOSAIC~\cite{dalgaty2024mosaic} replaces mesh-based neuromorphic routing with distributed memristor-based in-memory routers ($>$10$\times$ spike-routing energy reduction vs.\ SpiNNaker/Loihi digital mesh); Quarry~\cite{sun2021quarry} cuts ADC resolution from 7-bit to 3-bit with $\sim$32$\times$ energy reduction at $<$0.24\% accuracy loss using noise-aware training~\cite{joshi2020accurate}; adaptive memristor-based ADCs achieve 5-bit resolution with $\sim$15$\times$ energy and $\sim$13$\times$ area reduction versus conventional high-precision designs~\cite{hong2025adaptive_adc}. \emph{Spiking neural network encoding:} SNN-based processing replaces multi-bit ADCs with 1-bit comparators detecting spike presence~\cite{kim2024sae_snn}, particularly well-suited to inherently spike-like DVS data; dedicated ultra-low-power integrate-and-fire neuron circuits have been demonstrated in nanoscale CMOS~\cite{larsh2026spiking, siddique2024neuromorphic} and in side-contacted field-effect diode technology~\cite{motaman2025iff}, providing spiking readout blocks compatible with memristive SNN back-ends. \emph{Photonic direct coupling:} photomemristors accepting optical inputs bypass the DAC path entirely; the OEM array of~\cite{huang2025multimode} demonstrates this approach, with the output ADC remaining but the input DAC eliminated.

\begin{figure*}[!htbp]
\centering
\includegraphics[width=0.8\textwidth]{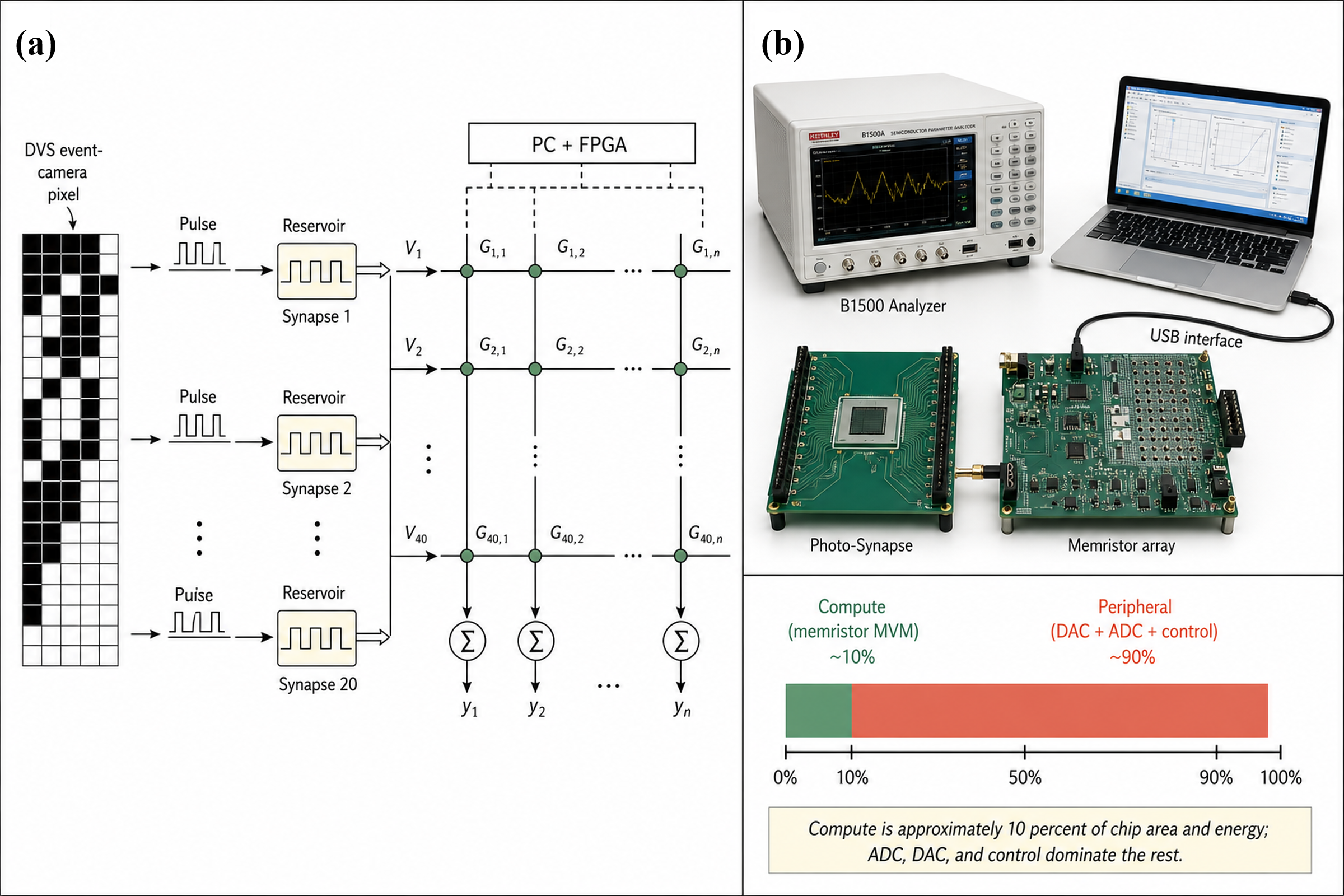}
\Description{Two-panel sensor-to-crossbar pipeline with a review-paper annotation. Panel (a): a reservoir-plus-crossbar back-end in which a 20-row DVS event-camera pixel array drives row-wise photo-synapses that feed a memristor array under PC and FPGA control, producing analog matrix-vector multiplication outputs y1=k(I1+ - I1-) ... yn=k(In+ - In-); the leftmost block is labelled "DVS event-camera pixel". Panel (b): a photograph of the fabricated hardware showing a B1500 analyser, the photo-synapse board, the memristor array, and the host workstation connected via USB. Bottom-right: a horizontal stacked-bar inset showing the area/energy split, ~10 percent compute (memristor MVM) in teal and ~90 percent peripheral (DAC + ADC + control) in red.}
\caption{End-to-end sensor-to-crossbar pipeline illustrating how an event-driven front-end couples to a memristor-array back-end. (a)~Reservoir-plus-crossbar back-end: a 20-row DVS event-camera pixel array drives row-wise photo-synapses that feed a memristor crossbar controlled by PC and FPGA; outputs $y_1,\dots,y_N$ accumulate analog MVM results. (b)~Fabricated hardware photograph (B1500 analyser, photo-synapse board, memristor array, host workstation). Bottom-right inset: area/energy budget split for the on-chip implementation, with the memristor MVM at ${\sim}10\%$ and DAC/ADC/control peripherals at ${\sim}90\%$ --- a representative split derived from the ISAAC budget breakdown~\cite{shafiee2016isaac} and the ADC-overhead analysis of Xia and Yang~\cite{xia2019memristive}, not measured on this specific chip. Adapted from Wang et al., \emph{npj Unconv.\ Comput.} \textbf{2}, 15 (2025)~\cite{wang2025bioinspired_opto} under CC BY 4.0. Every output line in panel~(a) requires an ADC conversion, and the host-software stack in panel~(b) is the consequence of that conversion cost.}
\label{fig:sensor_to_crossbar}
\end{figure*}

\begin{table*}[t]
\centering
\begin{threeparttable}
\caption{Comparison of Peripheral Interface Architectures for DVS-Memristor Integration, Ordered by Decreasing Overhead.}
\label{tab:adc_comparison}
\footnotesize
\setlength{\tabcolsep}{4pt}
\begin{tabular}{@{}>{\raggedright\arraybackslash}p{2.6cm} >{\raggedright\arraybackslash}p{3.7cm} >{\raggedright\arraybackslash}p{3.2cm} >{\raggedright\arraybackslash}p{3.0cm} c@{}}
\toprule
\textbf{Architecture} & \textbf{Description} & \textbf{Advantages} & \textbf{Limitations} & \textbf{DVS Fit} \\
\midrule
High-Precision ADC (8--12~bit)
  & Standard multi-bit analog-to-digital conversion of crossbar column currents.
  & Maximum numerical precision for complex inference.
  & Peripherals can dominate chip area and energy budgets in memristive CIM~\cite{shafiee2016isaac,xia2019memristive}; noise-limited ADC energy scales by $\sim$4$\times$ per additional bit~\cite{gholami2021survey}.
  & Low \\

Charge-Domain Converter
  & Switched-capacitor circuits integrate column current before conversion.
  & Mitigates nonlinear I-V characteristics of memristors~\cite{shafiee2016isaac}.
  & Large capacitors limit pixel-level integration density.
  & Moderate \\

Time-to-Digital (TDC)
  & Information encoded in pulse width or inter-spike interval.
  & Low power; naturally aligns with temporal nature of DVS events.
  & Sensitive to noise and jitter; complex time-domain logic.
  & High \\

1-Bit Sense Amplifier
  & Simple comparators determine threshold crossing (binary neural networks).
  & Minimal area/power; eliminates multi-bit ADC entirely.
  & Reduced inference accuracy for complex classification tasks.
  & High \\

ADC-less Spiking Interface
  & Direct pulse-modulation where DVS events gate memristor programming voltages.
  & Eliminates all data conversion; 10--66\% improvement vs.\ prior memristive SNN accelerators~\cite{kim2024sae_snn}.
  & Difficult to implement gradient-based learning rules.
  & Optimal \\
\bottomrule
\end{tabular}
\begin{tablenotes}
\small
\item Entries ordered from highest overhead (top) to lowest (bottom), corresponding to increasing DVS compatibility.
\item The ADC-less spiking interface achieves the full power reduction promise of neuromorphic DVS-memristor systems by operating entirely in the spike domain.
\end{tablenotes}
\end{threeparttable}
\end{table*}

A schematic of these architectures arranged on a single design axis appears in Supplement~S3; Figure~\ref{fig:sensor_to_crossbar} shows an end-to-end example of the same pipeline in a fabricated system, and Table~\ref{tab:adc_comparison} compares the mitigation strategies quantitatively.
The optimal solution depends on the application: high-accuracy tasks
(medical imaging, autonomous driving) may tolerate the area cost of
high-resolution ADCs, while energy-constrained edge deployments benefit
from SNN-compatible 1-bit readouts or photonic direct coupling.

\subsubsection{The Sparse-Event-to-Dense-Crossbar Mismatch}

DVS events are sparse (fewer than 5\% of pixels emit events in a given millisecond), while crossbar MVM activates all input rows simultaneously and incurs full ADC conversion on every column. Event-driven row activation, sub-array partitioning aligned to DVS regions, event-accumulation buffering, and column-parallel ADC gating are potential mitigations~\cite{kim2024sae_snn, hong2025adaptive_adc}; none has been demonstrated on a memristor-DVS workload. In practice, deployed pipelines pre-convert raw events to dense tensors (voxel grids~\cite{zhu2019voxelgrid}, time surfaces~\cite{lagorce2017hots}, EST/Event-Volume used by RVT~\cite{gehrig2023rvt}, GETransformer~\cite{peng2023get}, SSMs~\cite{zubic2024ssm}) before the crossbar input, mostly resolving the sparse-to-dense mismatch in software; the design question becomes which representation minimises accuracy loss at the shortest accumulation window. State-space models~\cite{zubic2024ssm} are structurally well matched to memristor crossbar acceleration (single MVM per hidden-state update, dense event-volume representation), but no published memristor-DVS implementation has mapped an SSM backbone to a crossbar. The research priority shifts from new crossbar circuits for raw-event processing to joint event-representation and memristor-precision co-design.

\section{Applications and Maturity Assessment}
\label{sec:applications}

The architectural analysis of Section~\ref{sec:architectures} described what has been built. This section asks what those builds deliver in application terms, with evidence grading as the analytical instrument. The framework appears below (Section~\ref{sec:evidence-framework}); the per-domain summaries that follow (Section~\ref{sec:app-analysis}) are its output. The chip-prototype reference design discussed in Section~\ref{sec:array-level} is the most complete integrated memristor+DVS system to date; the majority of remaining application claims rest on theoretical extrapolation, and Section~\ref{sec:gap-analysis} converts these into the engineering targets the roadmap addresses.

\subsection{Evidence-Graded Assessment Framework}
\label{sec:evidence-framework}

To separate measured results from speculation, this review classifies every application-relevant claim into three evidence levels:

\begin{itemize}
    \item \demonstrated{}: Fabricated hardware with measured performance data. The claim rests on a physical device or array tested under controlled conditions, with quantitative results reported.
    \item \simulated{}: Software simulation or circuit-level modeling. The claim is supported by numerical models but no physical device was fabricated for the specific claimed application.
    \item \projected{}: Theoretical extrapolation with no experimental data. The claim extends demonstrated device properties to an application domain without any simulation or measurement specific to that domain.
\end{itemize}

\paragraph{Grading rubric and boundary cases.}
Two boundary cases clarify the rubric. A device tested only with synthetic optical patterns is graded \demonstrated{} at the device level but not for a DVS application (event-stream constraints absent from static tests); a claim combining demonstrated DVS results from one paper with demonstrated memristor results from another is graded \projected{} because integration is unvalidated. SPICE studies using experimentally extracted parameters (e.g.,~\cite{vourkas2021inmemory}) are graded \demonstrated{} at device level and \simulated{} at system level. Grades are applied against this review's DVS-integration criterion (a fabricated, benchmarked memristor+DVS integrated system on standard event-camera data), not against the original authors' stated scope; a \projected{} grade reflects absence of integration evidence in the reviewed literature, not a shortfall in the original work.

Table~\ref{tab:evidence_matrix} maps 16 research topics to their highest evidence level, revealing that 7 of 16 remain at the \projected{} stage. Of the 146 papers in the corpus, ten define the memristor--DVS intersection at the device or array level~\cite{vourkas2021inmemory, laiho2015memristive, cao2023dual, cai2023insitu, wang2024perovskite, huang2025multimode, tan2023dynamic, li2018analogue, yoon2024memristor, zhou2023silk}: nine report fabricated hardware and one~\cite{yoon2024memristor} is a circuit-simulation study. None demonstrates a complete integrated memristor+DVS system tested on a real-world vision task. The remaining application claims combine DVS capabilities from one paper with memristor capabilities from another, a form of editorial extrapolation this review explicitly flags.

\begin{table*}[t]
\centering
\caption{Research Evidence Matrix for Memristor-DVS Integration. Each research topic is categorized by the highest level of evidence available.}
\label{tab:evidence_matrix}
\resizebox{\textwidth}{!}{%
\begin{threeparttable}
\small
\begin{tabular}{@{}l c c c l@{}}
\toprule
\textbf{Research Topic} & \textbf{\demonstrated} & \textbf{\simulated} & \textbf{\projected} & \textbf{Next Milestone} \\
\midrule
Photomemristor pixel (single device) & \checkmark & & & Array-scale uniformity \\
1D1M in-pixel memory & \checkmark & & & Endurance $>$10$^6$ cycles \\
Crossbar MVM (128$\times$64) & \checkmark & & & Scaling to megapixel \\
Multi-mode OEM (1k cells) & \checkmark & & & Scaling beyond 1k \\
Background subtraction (pixel) & \checkmark & & & Dynamic scene complexity \\
Retinomorphic reservoir & \checkmark & & & Frame-rate vs.\ $\mu$s DVS events \\
\midrule
DVS event processing on crossbar & & \checkmark & & Fabricated demonstration \\
CMOS-memristor hybrid power savings & & \checkmark & & Hardware validation \\
Point-cloud processing on memristor & & \checkmark & & DVS integration \\
\midrule
Complete memristor+DVS system & & & \checkmark & First integrated module \\
Memristor-DVS for autonomous driving & & & \checkmark & Vehicle-platform prototype \\
Memristor-DVS for AR/VR & & & \checkmark & Head-tracking prototype \\
Memristor-DVS for medical implant & & & \checkmark & Implantable prototype \\
Memristor-DVS for IoT deployment & & & \checkmark & Deployed module \\
On-chip learning with DVS events & & & \checkmark & DVS-specific learning demo \\
Wafer-scale memristor-DVS integration & & & \checkmark & Megapixel-scale fabrication \\
\bottomrule
\end{tabular}
\begin{tablenotes}
\small
\item Of 16 research topics, 6 have hardware demonstrations (device-level only), 3 have simulation evidence, and 7 remain at the projection stage.
\item The first complete system-level demonstration integrating memristors with a commercial DVS sensor would lift multiple rows simultaneously.
\end{tablenotes}
\end{threeparttable}%
}
\end{table*}

\subsection{Consolidated Application Analysis}
\label{sec:app-analysis}

Table~\ref{tab:applications} consolidates the six application domains into a single evidence-graded summary. The following paragraphs provide context for each domain.

\begin{table*}[t]
\centering
\caption{Consolidated Application Analysis of Memristor-Enhanced Dynamic Vision Systems. Evidence levels: \demonstrated{} = fabricated hardware with measured results; \simulated{} = software modeling; \projected{} = theoretical extrapolation.}
\label{tab:applications}
\resizebox{\textwidth}{!}{%
\begin{threeparttable}
\footnotesize
\setlength{\tabcolsep}{3pt}
\begin{tabular}{@{}l l l l l l@{}}
\toprule
\textbf{Domain} & \textbf{Ref.} & \textbf{Architecture} & \textbf{Key Result} & \textbf{Memristor Role} & \textbf{Evidence} \\
\midrule
\multirow{3}{*}{Robotics}
  & \cite{falanga2020dynamic} & CMOS DVS (no memristor) & 3.5~ms obstacle avoidance & None (baseline) & \demonstrated \\
  & \cite{tan2023dynamic} & Photomemristor reservoir & Real-time trajectory prediction & R1 & \demonstrated \\
  & \cite{yoon2024memristor} & CMOS-memristor hybrid & 75--79\% power reduction & R3 & \simulated \\
\midrule
\multirow{3}{*}{Automotive}
  & \cite{huang2025multimode} & Multi-mode OEM & 20$\times$ energy reduction$^\dagger$ & R1/R3 & \demonstrated \\
  & \cite{wang2024memristorbased} & 40$\times$25 memristor array & 94\% accuracy, 10 environments & R3 & \simulated \\
  & \cite{yoon2024memristor} & Hybrid DVS processor & 75--79\% power saved & R3 & \simulated \\
\midrule
\multirow{2}{*}{AR/VR}
  & \cite{tan2023dynamic} & Photomemristor reservoir & Motion prediction ($<$20~ms) & R1 & \demonstrated$^*$ \\
  & --- & --- & Head-tracking / gesture / gaze demonstrator & --- & \projected \\
\midrule
\multirow{3}{*}{Surveillance}
  & \cite{laiho2015memristive} & TaOx pixel array & Sub-ms background subtraction & R2 & \demonstrated \\
  & \cite{huang2025multimode} & Multi-mode OEM & 91--96\% accuracy & R1/R3 & \demonstrated \\
  & \cite{yoon2024memristor} & Hybrid CMOS & 75\% power savings & R3 & \simulated \\
\midrule
\multirow{2}{*}{Medical}
  & \cite{zhou2023silk} & Silk-protein dual-mode & Retina-to-cortex analog & R1/R3 & \demonstrated \\
  & --- & --- & Implantable retinal / surgical demonstrator & --- & \projected \\
\midrule
\multirow{2}{*}{IoT/Edge}
  & \cite{lao2022ultralow} & Self-powered sensor & Energy-harvesting vision & R1 & \demonstrated \\
  & --- & --- & Deployed IoT memristor-DVS module & --- & \projected \\
\bottomrule
\end{tabular}
\begin{tablenotes}
\small
\item $^*$ Device demonstrated for general motion prediction, not specifically in an AR/VR system.
\item $^\dagger$ GPU-relative energy comparison uncontrolled for model complexity (single-layer OEM versus deeper GPU network).
\item Of 15 application claims, 8 are \demonstrated{} at device level, 4 are \simulated{}, and 3 are \projected{}.
\item The first complete integrated memristor+DVS system deployed in any application domain would lift its row from \projected{} to \demonstrated{}.
\end{tablenotes}
\end{threeparttable}%
}
\end{table*}

\subsubsection{Robotics and industrial automation.}
The strongest robotics claim involves DVS-based obstacle avoidance at 3.5~ms overall algorithm latency, enabling quadrotor avoidance of thrown objects approaching at relative closure speeds up to 10~m/s~\cite{falanga2020dynamic} (not sustained flight speeds); the latency budget itself derives from the sense-and-avoid analysis of Falanga et al.~\cite{falanga2019howfast}. However, this result uses a pure CMOS DVS sensor with no memristor component (\demonstrated{} for DVS, not applicable for memristors). Separately, retinomorphic photomemristor arrays have demonstrated motion recognition and trajectory prediction at up to 60~Hz with ON/OFF ratios on the order of $10^2$~\cite{tan2023dynamic} (\demonstrated{} at device level). A CMOS-memristor hybrid pipeline reported 75--79\% power savings with 0.5--0.75\% accuracy loss on the gesture/digit DVS benchmarks POKER-DVS and MNIST-DVS~\cite{yoon2024memristor}, but the result comes from circuit simulation rather than fabricated hardware, and whether the savings transfer to event streams collected in a robotic manipulation or navigation setting has not been validated (\simulated{}). No study has demonstrated a memristor-integrated DVS system performing robotic manipulation or navigation in a real environment.

\subsubsection{Autonomous vehicles and smart transportation.}
The multi-mode optoelectronic memristor (OEM) array achieved 20$\times$ energy reduction compared to GPU-based processing across tasks including object tracking and motion prediction~\cite{huang2025multimode} (\demonstrated{} at the 128$\times$8 array level). A 40$\times$25 memristor array processed differentially encoded visual motion cues, reaching 94\% accuracy across ten road environments~\cite{wang2024memristorbased} (\simulated{}: the accuracy was measured in simulation, not in a vehicle). The 75--79\% CMOS-memristor power-saving figure of~\cite{yoon2024memristor} is \simulated{} on POKER-DVS and MNIST-DVS rather than on automotive event-camera data, leaving its automotive transfer unvalidated.

Claims of microsecond obstacle detection, adaptive driving style learning, and real-time 3D point-cloud processing on memristor arrays are \projected{}. No memristor-DVS system has been tested on a vehicle platform or on standard automotive event-camera datasets (DSEC~\cite{gehrig2021dsec}, Gen1~\cite{detournemire2020gen1}, eTraM~\cite{verma2024etram}). Event-based automotive perception has matured rapidly: Gehrig and Scaramuzza~\cite{gehrig2024lowlatency} established that event+RGB camera fusion matches 5{,}000~fps camera detection latency at 45~fps bandwidth, and the current state of event-based detection backbones (Recurrent Vision Transformers~\cite{gehrig2023rvt}, Group Event Transformer~\cite{peng2023get}, state-space models~\cite{zubic2024ssm}, hybrid SNNs~\cite{ahmed2025hybrid_snn}) defines the digital baseline against which any memristor-DVS automotive proposal must compete; none has been mapped to a memristor crossbar.

\subsubsection{Augmented and virtual reality.}
AR/VR is an open application track for memristor-DVS integration: head tracking, gesture recognition, and gaze estimation are all \projected{} at this stage. The sub-20~ms motion-to-photon requirement for comfortable VR is well-matched to memristor crossbar latency; hydrogel photonic memristors~\cite{wang2025hydrogel} and flexible neuromorphic arrays~\cite{su2024nonvon} provide device building blocks, awaiting integration with DVS event streams.

\subsubsection{Intelligent surveillance and security.}
Oxide-based memristive pixels demonstrated per-pixel background subtraction with sub-millisecond update times~\cite{laiho2015memristive} (\demonstrated{} at device level), the closest any memristor work comes to a surveillance-relevant function, on a small device count. The 75--79\% power-saving result~\cite{yoon2024memristor} (\simulated{}) and the reservoir-encoder accuracy of~\cite{tan2023dynamic} (\demonstrated{} at array level) are on gesture/digit benchmarks; surveillance footage at camera scale is the natural next demonstrator. Distributed intelligence, proactive threat detection, and privacy-preserving local processing are \projected{} application targets that follow from those device-level results.

\subsubsection{Medical and biomedical applications.}
Silk-protein dual-mode devices span R1+R3 roles on a biocompatible substrate~\cite{zhou2023silk} (\demonstrated{} at device level), processing optical input directly. Memristor-based retinal prosthetics, surgical vision systems, and wearable health monitors with DVS input are \projected{} application targets where the biocompatibility of silk and the pixel-level density of photomemristors meet a clear deployment surface.

\subsubsection{IoT and edge computing.}
A self-powered perovskite reservoir coupled to a photovoltaic layer demonstrated near-zero standby power on a curated 4-class vehicle-flow DVS dataset~\cite{lao2022ultralow} (\demonstrated{} at small-array level; 100\% reported accuracy on a 4-class benchmark should not be read as practical recognition performance). Full-scale IoT applications (smart buildings, predictive maintenance, environmental surveillance) remain \projected{} demonstrator targets that build directly on this self-powered architecture.

\subsubsection{The digital neuromorphic baseline.}
\label{sec:digital-baseline}
The double baseline of contribution~C3 names the digital neuromorphic processors against which any memristor-DVS system must be evaluated. SynSense's Speck~\cite{richter2023speck} is the only commercial event-sensor-plus-neural-processor product approximating the monolithic integration envisioned for memristor-DVS, integrating a DVS pixel array with a 327K-neuron on-chip convolutional SNN accelerator at sub-millisecond per-event latency. Intel Loihi~2~\cite{orchard2021loihi2} and SpiNNaker~2~\cite{gonzalez2024spinnaker2} provide multi-core alternatives at low-power operation; neuromorphic robust-estimation~\cite{ahmadvand2025robust} and cloud-to-edge event-inference frameworks~\cite{ahmadvand2024cloudedge} illustrate the algorithm spectrum these digital platforms host without memristor mapping. The \emph{integrated} memristor-DVS target is therefore well-defined: surpass Speck's measured power--latency--accuracy envelope on a standard event-camera benchmark (DVS128 Gesture, N-Cars, Gen1, or DSEC) while delivering additional benefit in density or on-chip learning. The case for memristor integration rests on substantial power reduction at iso-accuracy or on capabilities (multi-level analog storage, on-chip learning) that digital neuromorphic processors do not natively provide; merely matching Loihi~2 or Speck at similar wattage does not justify the integration complexity, analog non-idealities, and endurance exposure of a memristor crossbar.

\subsubsection{Cross-domain observations.}
The same ten core references anchor all six application domains. DVS-specific results (obstacle avoidance latency, tracking speed) come from pure-CMOS DVS sensors, while memristor-specific results (power savings, in-sensor computing) come from devices that do not yet interface with DVS cameras. The first integrated demonstrator that bridges the two would convert system-level claims from \projected{} to \demonstrated{} across multiple domains simultaneously --- the structural reason why Milestone~N1 (Section~\ref{sec:roadmap}) is the highest-impact near-term target.

\subsection{Technology Readiness as a Secondary View}
\label{sec:trl}

\begin{figure}[!htbp]
    \centering
    \includegraphics[width=\columnwidth]{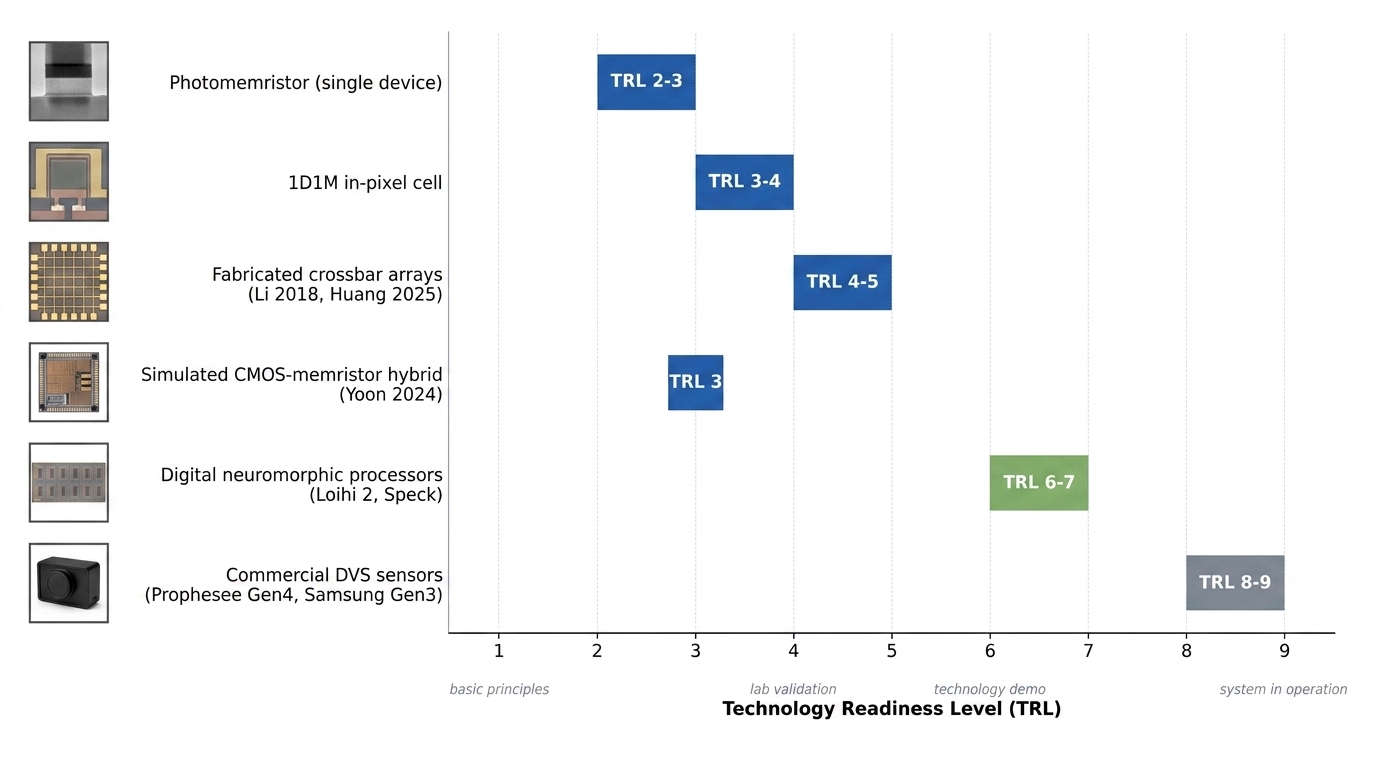}
    \Description{Horizontal-bar chart positioning each implementation category on the 9-level Technology Readiness Level scale. Bars (top to bottom): photomemristor single device at TRL 2-3, 1D1M in-pixel cell at TRL 3-4, fabricated crossbar arrays (Li 2018, Huang 2025) at TRL 4-5, simulated CMOS-memristor hybrid (Yoon 2024) at TRL 3 (all in primary teal-blue); digital neuromorphic processors (Loihi 2, Speck) at TRL 6-7 in pale sage; commercial DVS sensors (Prophesee Gen4, Samsung Gen3) at TRL 8-9 in muted slate. The x-axis runs from TRL 1 to 9; no Valley-of-Death overlay.}
    \caption{Technology Readiness Level (TRL) assessment. Horizontal bars indicate the assessed TRL range for each implementation category. No memristor-DVS implementation has reached the validated-systems regime that commercial DVS sensors and digital neuromorphic processors already occupy; these reference bars are shown for context.}
    \label{fig:trl}
\end{figure}

The evidence-graded matrix in Table~\ref{tab:evidence_matrix} carries the primary maturity assessment. As a secondary view, Figure~\ref{fig:trl} maps the reviewed implementations onto the nine-level Technology Readiness Level (TRL) scale: photomemristor single devices sit at TRL 2--3, 1D1M pixels at TRL 3--4, fabricated crossbar arrays~\cite{li2018analogue, huang2025multimode} at TRL 4--5, and the simulation-only CMOS--memristor hybrid~\cite{yoon2024memristor} at TRL 3. No memristor-DVS implementation crosses TRL~5 for a vision application; commercial DVS sensors (Prophesee Gen4, Samsung Gen3) sit at TRL~8--9 and digital neuromorphic processors (Loihi~2, Speck) at TRL~6--7. The gap between the memristor integration layer and the surrounding ecosystem is therefore at least three TRL levels.

\subsection{Five Engineering Targets}
\label{sec:gap-analysis}

Five engineering targets define the path from the current research state to practical deployment, each grounded in the evidence-graded analysis above.

\textbf{Target~1: First integrated system.} The first integration milestone is a complete memristor+DVS module with the DVS pixel array and the memristor processing layer on the same chip or in the same package. Reaching this milestone moves the System Integration row of Table~\ref{tab:evidence_matrix} from \projected{} to \demonstrated{}.

\textbf{Target~2: Real-world deployment.} Field validation under variable illumination, temperature drift, and vibration is the second maturity gate. Characterising memristor robustness under continuous, uncontrolled event streams converts laboratory benchmarks into operational evidence.

\textbf{Target~3: Evidence in every application domain.} Three application domains (AR/VR, medical, IoT) currently sit at the \projected{} tier; the next demonstrator in each lifts that domain into the \demonstrated{} tier and broadens the integration's deployment surface.

\textbf{Target~4: Endurance margin for continuous event traffic.} A DVS sensor at $10^5$--$10^7$ events per second per pixel cluster generates write traffic that exhausts a $10^6$-endurance device in seconds. Most reviewed devices fall in the $10^6$--$10^9$ range; closing the margin to the $10^{10}$ N2 criterion (Section~\ref{sec:roadmap}) opens the door to always-on workloads.

\textbf{Target~5: Community memristor-DVS benchmark.} A standardized benchmark coupling event-camera datasets with memristor device models would enable cross-study comparison and measurable progress. NeuroBench~\cite{yik2025neurobench} provides the algorithmic-and-systems infrastructure; the B1--B6 specification of Section~\ref{sec:standardization} adds the analog memristor-in-the-loop layer.

The five targets are uniform across the application matrix: each domain reaches deployment-grade integration when targets 1--5 are met. The chip-level demonstrators that already approach target 1 (e.g., the Chang/Lele SoC~\cite{sharma2023isscc, sharma2024jssc}, which reaches laboratory-validated target-tracking benchmarks) define the trajectory; the milestones in Section~\ref{sec:roadmap} make that trajectory measurable.

\section{Challenges, Open Problems, and Research Roadmap}
\label{sec:challenges}

Memristor-DVS \emph{integration} spans TRL~2--5 even as memristor compute-in-memory has reached TRL~6--7 at industrial scale~\cite{ambrogio2023analogai, wan2022neurram, rao2023thousands} and digital neuromorphic processors running DVS streams (Loihi~2, SynSense Speck) are commercially shipping. This section identifies the specific technical barriers between the present state and the integrated system the field is targeting, organised from device level through system integration to manufacturing and standardisation, and converts each barrier into a falsifiable roadmap milestone in Section~\ref{sec:roadmap}. Granular per-barrier deep-dives appear in Supplement~S2.

\subsection{Device-Level Barriers: Variability and Endurance}
\label{sec:device-challenges}

D2D variability spans a wide range by material system: SiN\textsubscript{x} 1D1M pixels report sub-percent SPICE-level variability across simulated multi-level conductance states~\cite{vourkas2021inmemory} and perovskite ferroelectrics show high linearity~\cite{wang2024perovskite}, while solution-processed MoS\textsubscript{2} arrays report 19.7\% (set) and 18.5\% (reset) D2D variability~\cite{tang2022waferscale}. No material system currently satisfies both $<$5\% D2D and ON/OFF $>$100 at array scale: filamentary devices (HfO\textsubscript{x}, TaO\textsubscript{x}) achieve large dynamic range with high D2D; interfacial devices (ferroelectric, charge-trapping) achieve uniformity at the cost of dynamic range. C2C variability accumulates over millions of writes and is uncharacterised under DVS-like bursty traffic; metal oxides show 3--10\% C2C spread, perovskite ferroelectrics approach unity linearity~\cite{wang2024perovskite} (deeper analysis in Supplement~S2).

Endurance is the most application-binding constraint. A DVS sensor at $10^5$--$10^7$ events per second per pixel cluster generates write traffic that exhausts a $10^6$-endurance device in seconds, yet most reviewed devices fall in the $10^6$--$10^9$ range (Table~\ref{tab:performance}). Perovskite dual-functional photomemristors~\cite{cao2023dual} describe only qualitative cyclic stability; perovskite ferroelectrics~\cite{wang2024perovskite} claim $>$10$^9$ cycles, but not validated under sustained DVS-like write patterns. The endurance burden is role-dependent: R1 (photomemristor) is worst since every photon-induced change is a write; R3 (crossbar) is least constrained as weights update infrequently; R2 falls between. The fast-switching/long-retention trade-off compounds this: perovskite photomemristors retain state $>$7000\,s~\cite{cao2023dual}, adequate for temporal integration but insufficient for always-on driving.

\subsection{Integration Barriers: CMOS, Sneak Paths, ADCs}
\label{sec:integration-challenges}

BEOL thermal budget ($<$400\si{\degreeCelsius}) is satisfied by metal-oxide memristors (HfO\textsubscript{x}, TaO\textsubscript{x})~\cite{li2018analogue}; perovskite and organic devices require $<$150\si{\degreeCelsius}, compatible with BEOL but limiting annealing steps that improve uniformity. Wafer-scale passive crossbar fabrication is demonstrated~\cite{choi2025waferscale} but at array dimensions far below the megapixel scale required for DVS integration.

Sneak-path currents in passive crossbars corrupt read/write operations. The 1T1R topology eliminates them at the cost of density and 3D stacking; a 32$\times$32 1S1R crossbar with HfSe\textsubscript{2} memristors and Si diode selectors achieved 89\% yield over 992 devices and suppressed sneak paths sufficiently for MVM~\cite{jain2025heterogeneous}. For DVS integration where high resolution demands large dense arrays, the 1S1R path is preferred but requires selector $>$10$^4$ nonlinearity, restricting material choices (full selector trade-off analysis in Supplement~S2). Wire resistance in large arrays causes IR drop degrading weight precision~\cite{hu2016dotproduct}, placing a practical upper bound of $\sim$256--512 rows; a 1280$\times$720 DVS sensor would require partitioning into hundreds of sub-arrays.

The peripheral ADC/DAC overhead (analyzed in full in Section~\ref{sec:adc-bottleneck}) is the single largest barrier to energy-efficient integration: peripheral circuits can dominate both chip area and energy, ADC energy scales $\sim$$4\times$ per additional bit~\cite{shafiee2016isaac, xia2019memristive}, and DVS microsecond resolution demands fast conversion. Mitigation strategies (1-bit spiking interfaces~\cite{kim2024sae_snn}, adaptive ADCs~\cite{hong2025adaptive_adc}, photonic direct coupling) reduce but do not yet eliminate this overhead.

\subsection{Manufacturing and Scalability Barriers}
\label{sec:manufacturing}

Scaling memristor arrays from laboratory ($<$1{,}000 cells) to production scale (megapixel) demands wafer-level uniformity that current processes do not deliver for most material systems. Solution-processed MoS\textsubscript{2} arrays achieved $1 \times 10^7$ cycle endurance with 19.7\% D2D variability~\cite{tang2022waferscale}; Choi et al.~\cite{choi2025waferscale} scaled passive memristive crossbar fabrication to a 4-inch wafer at $>$95\% yield, demonstrating CMOS-compatible wafer-scale processing but at array dimensions still far below megapixel. At 99.8\% yield~\cite{li2018analogue}, a 1280$\times$720 array would still contain $\sim$1{,}843 defective devices; fault types accumulate over lifetime rather than appearing at fabrication, and fault-tolerance approaches~\cite{liu2017rescuing, feinberg2018making, yousuf2025ensemble, shruthi2026fault_survey} have not been characterised under DVS-like bursty write traffic where per-pixel activity stresses cells unequally.

Industrial readiness has begun: Weebit Nano's AEC-Q100 automotive qualification of a 1T1R ReRAM~\cite{weebit2025aecq100} is the first commercial memristor to pass automotive stress testing, but at $10^5$ qualified endurance, four orders below research targets~\cite{intel2022optane}. HfO\textsubscript{2} BEOL stacks add minimally to foundry cost~\cite{martemucci2025ferroelectric}; exotic stacks (perovskites, organics, 2D films) require non-standard equipment. Initial weight programming consumes $10^5$--$10^6$ write cycles per cell via iterative write-verify~\cite{song2024precision}, a non-trivial fraction of the endurance budget; error-aware probabilistic training reduces subsequent write-cycle cost to $<$0.1\% of conventional~\cite{liu2025error_aware}.

On-chip learning for real-time DVS processing~\cite{ambrogio2018equivalent, rao2023thousands, ahmadvand2024cloudedge}, monolithic 3D thermal management~\cite{li2024_3d_multilayered}, long-term reliability under DVS aging~\cite{cao2023dual, zhao2026hightemp}, and robustness under adverse weather conditions are open problems with active research lines; their detailed analyses appear in Supplement~S2.

\subsection{Standardization and Benchmarking}
\label{sec:standardization}

NeuroBench~\cite{yik2025neurobench} provides standardized benchmarks for neuromorphic computing with Algorithm and System tracks, but neither track includes memristor-in-the-loop evaluation or analog device non-ideality modeling. JEDEC efforts address digital memory interfaces, not analog crossbar operation~\cite{jedec2025ddr5}; SNABSuite~\cite{ostrau2022benchmarking} targets digital neuromorphic processors. The field needs benchmarks that couple event-camera datasets with memristor device models. The following minimum specification, designed to be implementable today on top of NeuroBench-style infrastructure, is proposed for any community memristor--DVS benchmark:

\begin{itemize}
    \item \textbf{(B1)} A fixed event-camera test stream from DVS128-Gesture, Gen1~\cite{detournemire2020gen1}, DSEC~\cite{gehrig2021dsec}, or eTraM~\cite{verma2024etram}, with canonical resolution and train/val/test split.
    \item \textbf{(B2)} A specified dense-tensor event representation (voxel grid, time surface, EST/SBT) and accumulation window; alternative representations reported as ablations.
    \item \textbf{(B3)} A SPICE- or PyTorch-callable device model with stated C2C/D2D variability, bit precision, endurance budget, and read noise; an idealised variant (upper bound) and a published-parameter HfO\textsubscript{x}/PCM variant (realistic estimate).
    \item \textbf{(B4)} An open-source crossbar simulator accounting for sneak paths and IR drop at baseline 256$\times$256 array size~\cite{rao2023thousands, wan2022neurram}.
    \item \textbf{(B5)} A joint primary score combining accuracy, energy/inference (J/event at the specified window), latency, and projected device lifetime; a Loihi~2 or Speck reference reported alongside as the digital baseline any memristor entry must beat.
    \item \textbf{(B6)} Public code, parameter files, and an environment lock-file; vendor-only datasheet entries excluded.
\end{itemize}

Without such a specification, claims of ``20$\times$ energy savings''~\cite{huang2025multimode} or ``75--79\% power reduction''~\cite{yoon2024memristor} cannot be compared to each other or to digital neuromorphic baselines on equivalent tasks. The ten core intersection papers~\cite{vourkas2021inmemory, laiho2015memristive, cao2023dual, cai2023insitu, wang2024perovskite, huang2025multimode, tan2023dynamic, li2018analogue, yoon2024memristor, zhou2023silk} report device-level results without publicly callable device models, simulation scripts, or raw measurement data (Supplement~S2); B1--B6 closes that gap by giving the community a common interface to implement. Memristor-DVS work spans three analytical layers (device-level circuit simulation, array-level non-ideality modelling, system-level event-driven inference), each with its own tooling ecosystem (Table~\ref{tab:tools}); few studies couple all three in a single co-simulation loop.

\begin{table*}[t]
\centering
\begin{threeparttable}
\caption{Simulation, modelling, and implementation tools cited in the memristor-DVS corpus and adjacent literature. Coverage spans the device level (single-cell models), the array level (crossbar non-idealities, ADC peripherals), and the system level (event-driven SNN training and inference). Columns indicate the analytical layer the tool primarily targets and an example use in the reviewed corpus.}
\label{tab:tools}
\footnotesize
\setlength{\tabcolsep}{3pt}
\begin{tabular}{@{}p{2.6cm} p{2.6cm} p{3.0cm} p{5.4cm}@{}}
\toprule
\textbf{Tool} & \textbf{Type} & \textbf{Primary scope} & \textbf{Example use in corpus} \\
\midrule
SPICE (HSPICE, ngspice)
  & Commercial / open-source
  & Device + small-array circuit
  & SiN\textsubscript{x} 1D1M pixel study~\cite{vourkas2021inmemory} \\

Cadence Spectre / Virtuoso
  & Commercial
  & Mixed-signal full-chip
  & Hybrid CMOS--memristor simulation~\cite{yoon2024memristor} \\

TEAM model
  & Open-source (Verilog-A)
  & Threshold-adaptive memristor compact model
  & Standard compact model~\cite{kvatinsky2013team} \\

NeuroSim~\cite{chen2018neurosim}
  & Open-source
  & Crossbar + ADC + system-level metrics
  & Macro-modelling for memristor accelerators \\

IBM AIHWKit
  & Open-source (PyTorch)
  & Analog DNN training under non-idealities
  & Noise-aware weight training, PCM transfer~\cite{joshi2020accurate} \\

snnTorch / Norse / Brian2
  & Open-source (Python)
  & SNN training and event-stream inference
  & Spiking detectors on event datasets~\cite{cordone2022object, ahmed2025hybrid_snn} \\

PyTorch / TF + event front-end
  & Open-source
  & End-to-end event-vision DL
  & Event-representation backbones~\cite{gehrig2023rvt, peng2023get, zubic2024ssm} \\

OpenAlex / Crossref / arXiv APIs
  & Open service
  & Literature corpus construction
  & PRISMA screening pipeline (Section~\ref{sec:methodology}) \\
\bottomrule
\end{tabular}
\begin{tablenotes}
\small
\item Most reviewed memristor-DVS device papers use SPICE-class tools for single-cell or small-array characterisation and Python deep-learning frameworks for the algorithmic side; few studies couple the two layers in a single co-simulation. The release of public tooling that closes this gap (memristor-aware extensions to AIHWKit, NeuroSim integration with event-vision frontends) is a standardisation milestone called out in Section~\ref{sec:standardization}.
\end{tablenotes}
\end{threeparttable}
\end{table*}

\subsection{Research Roadmap}
\label{sec:roadmap}

The following roadmap states specific, falsifiable milestones rather than vague expectations. Each milestone is stated as a target to be achieved, not a prediction of what will occur.

\subsubsection{Near-term (2025--2027): Mid-scale prototypes with DVS event processing.}
The immediate priority is to demonstrate memristor arrays processing real DVS event streams, closing the gap between the two independently demonstrated technologies.

\begin{itemize}
    \item \textbf{Milestone~N1 (component-level):} A memristor crossbar of at least $256\times256$ cells (matching the fabrication scale of current memristor-CIM SOTA~\cite{rao2023thousands, wan2022neurram}) ingests a pre-recorded DVS event tensor from a standard benchmark (DVS128~Gesture or N-Cars) and reports measured per-inference energy and latency directly on memristor hardware, rather than relying on the SRAM near-memory path used in the Chang/Lele 2023/2024 hybrid SoC~\cite{sharma2023isscc, sharma2024jssc}. Target: match the measured energy/latency envelope of digital neuromorphic processors (Loihi~2, Speck) on the same workload at iso-accuracy. The end-to-end integrated system milestone ($\geq$90\% accuracy on DVS128~Gesture at $\leq$10~mW system power and $\leq$5~ms latency on a coupled DVS--memristor pipeline) is the more ambitious target that follows once~N1 is met.
    \item \textbf{Milestone~N2:} Device-level endurance exceeds $10^{10}$ cycles for at least one CMOS-compatible material system (metal oxide or 2D material), validated under DVS-like write patterns (bursty, variable-rate).
    \item \textbf{Milestone~N3:} A standardized memristor-DVS benchmark is published, specifying device models, event-camera datasets, and joint accuracy/power/lifetime metrics. Community adoption by at least three independent research groups.
\end{itemize}

Each near-term milestone closes a specific deficit identified earlier: \textbf{N1} closes Gap~1 (no integrated system) and forces a \demonstrated{} entry into Table~\ref{tab:evidence_matrix} for the System Integration row currently graded \projected{}; \textbf{N2} closes Gap~4 (endurance mismatch); \textbf{N3} closes Gap~5 (no standardised benchmarks) and instantiates the B1--B6 specification of Section~\ref{sec:standardization}. Milestone~N1 is achievable using Role~3 integration (external crossbar processing DVS events) without requiring monolithic integration.

\paragraph{How the N1 target is calibrated.} The 90\%/10\,mW/5\,ms triple matches the iso-accuracy envelope of shipping baselines: Speck~\cite{richter2023speck} achieves mid-90\% DVS128~Gesture accuracy at sub-millisecond latency, and the GenX320 draws $\sim$3\,mW~\cite{prophesee2024genx320}. The design priority is joint event-representation/memristor-precision co-design: which voxel-grid or time-surface format minimises accuracy loss at the lowest sustainable analog precision.

\subsubsection{Mid-term (2027--2030): Monolithic integration and system-level validation.}
The mid-term goal is a physically integrated memristor-DVS module validated on realistic vision tasks.

\begin{itemize}
    \item \textbf{Milestone~M1:} A monolithic 3D-integrated chip combines a DVS pixel array ($\geq$640$\times$480 resolution, matching modern automotive-grade event sensors) with a memristor crossbar processing layer. The module achieves $\geq$85\% accuracy on a standard event-camera detection benchmark (Gen1, DSEC, or eTraM) at $\leq$50~mW total power, undercutting the measured Speck + external DVS pipeline on the same task. The 50\,mW envelope derives from current commercial VGA-resolution event sensors operating in the few-hundred-mW envelope (Table~\ref{tab:dvs_sensors}) plus a digital backend in the 100s of mW range; an integrated memristor backend that delivers the same accuracy at $\leq$50\,mW total represents a $\geq$8$\times$ system-level reduction, the threshold at which memristor integration justifies its design and reliability cost.
    \item \textbf{Milestone~M2:} D2D variability below 5\% achieved at wafer scale ($\geq$200~mm) for at least one material system, with $>$99\% array yield for arrays exceeding $10^4$ cells.
    \item \textbf{Milestone~M3:} ADC-less or sub-4-bit ADC interface architectures demonstrate $<$2\% accuracy loss compared to full-precision digital baselines on event-camera classification tasks, with measured (not simulated) power consumption.
\end{itemize}

Each mid-term milestone closes a further deficit: \textbf{M1} forces \demonstrated{} entries into Table~\ref{tab:evidence_matrix} for Monolithic Integration and Real-World Deployment (Gaps~1, 2); \textbf{M2} closes the wafer-scale uniformity gap blocking megapixel arrays; \textbf{M3} closes the ADC peripheral bottleneck quantified in Section~\ref{sec:adc-bottleneck}. The 2027--2030 window is a falsifiable target, not a forecast: if the field does not meet M1 within the window, the roadmap has been falsified at that gate.

\subsubsection{Long-term (2030+): Deployment in safety-critical and commercial applications.}
The long-term goal is field deployment with certified reliability.

\begin{itemize}
    \item \textbf{Milestone~L1:} A memristor-DVS module deployed in a safety-critical application (automotive perception or medical device) with reliability certification (e.g., AEC-Q100 for automotive, IEC~62304 for medical). Endurance exceeds $10^{12}$ cycles. Mean time between failures exceeds 10{,}000 hours of continuous operation.
    \item \textbf{Milestone~L2:} Memristor-DVS systems with embedded on-chip learning achieve continuous adaptation to changing environments (day/night, weather, aging) without external retraining, validated over $>$1{,}000 hours of field operation.
    \item \textbf{Milestone~L3:} A commercially available memristor-DVS module at cost parity ($<$2$\times$) with equivalent-capability digital neuromorphic solutions (e.g., Loihi~2 + DVS camera).
\end{itemize}

\begin{figure*}[!htbp]
    \centering
    \includegraphics[width=0.85\textwidth]{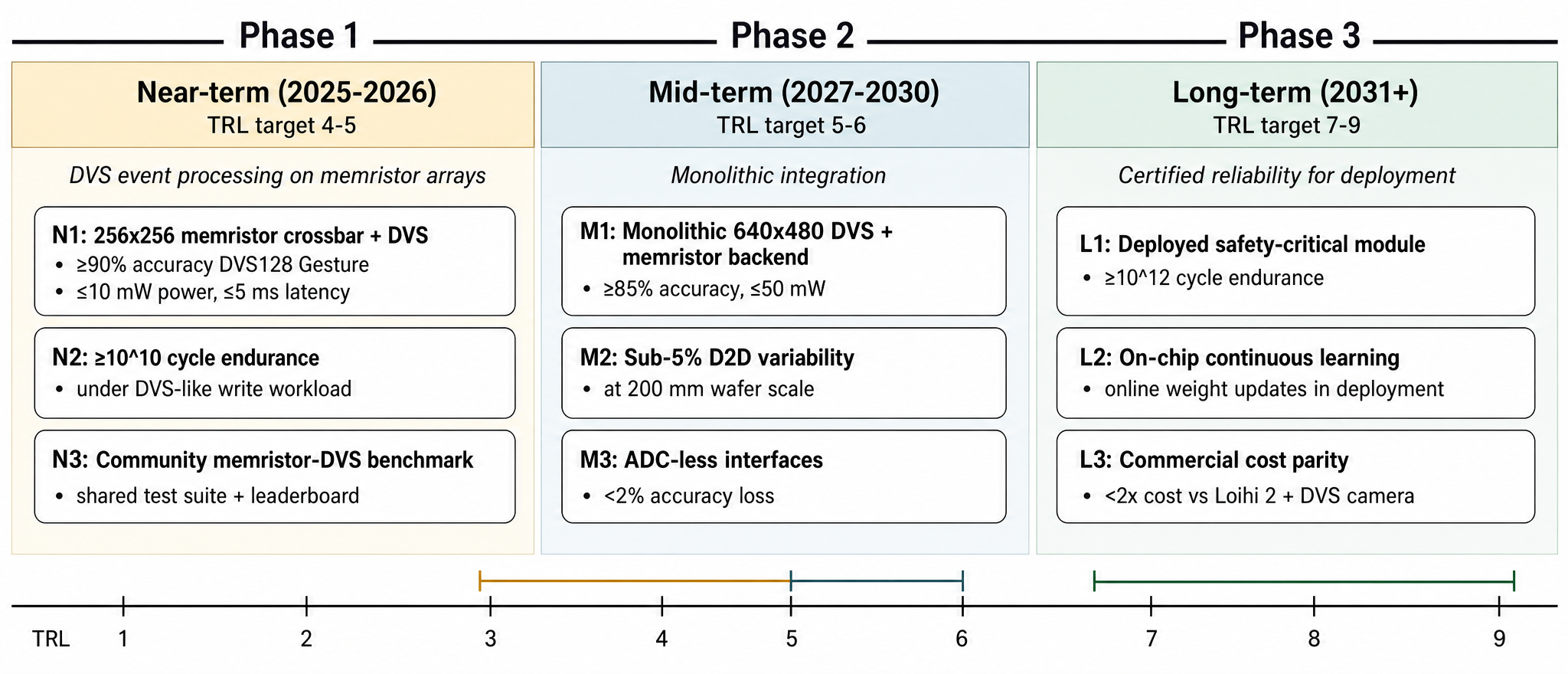}
    \Description{Three-phase horizontal roadmap. Near-term (2025-2027) lists milestones N1 (256-by-256 memristor crossbar coupled to DVS, 90 percent accuracy on DVS128 Gesture at 10 milliwatts, 5 ms latency), N2 (ten-to-the-tenth cycle endurance under DVS-like writes), and N3 (community memristor-DVS benchmark). Mid-term (2027-2030): M1 (monolithic 640-by-480 DVS plus memristor backend at 85 percent accuracy, 50 milliwatts), M2 (sub-5 percent D2D variability at 200 mm wafer scale), M3 (ADC-less interfaces with under 2 percent accuracy loss). Long-term (2030+): L1 (deployed safety-critical module with ten-to-the-twelfth cycles), L2 (on-chip continuous learning), L3 (commercial cost parity with digital neuromorphic plus DVS combinations).}
    \caption{Research roadmap for memristor-DVS integration. Three phases with falsifiable milestones. Current state: TRL~2--5. Near-term targets focus on DVS event processing on memristor arrays. Mid-term targets require monolithic integration. Long-term targets require certified reliability for deployment.}
    \label{fig:roadmap}
\end{figure*}

Each long-term milestone closes the remaining application-deployment deficit: \textbf{L1} forces a \demonstrated{} entry into Table~\ref{tab:evidence_matrix} for safety-critical deployment and lifts at least one application-domain row of Table~\ref{tab:applications} out of the \projected{} grade; \textbf{L2} closes the on-chip-learning deficit; \textbf{L3} closes the cost-parity gap that determines commercial viability against the digital-neuromorphic baseline. Figure~\ref{fig:roadmap} summarizes the three-phase roadmap with concrete milestones.

\section{Conclusion}
\label{sec:conclusion}

This review examined the integration of memristor technologies with dynamic vision sensors through 146 publications screened from 28{,}252 records, guided by three research questions. The answers follow.

\textbf{RQ1: Architectural paradigms and trade-offs.}
Three integration paradigms emerge from the reviewed literature. Pixel-level integration (Role~1: photomemristors; Role~2: 1D1M in-pixel memory) co-locates sensing and memory within each pixel, achieving the lowest projected power consumption, exemplified by perovskite photomemristors operating at $3 \times 10^{-11}$~W~\cite{cao2023dual} and SiN\textsubscript{x} 1D1M pixels with eight SPICE-distinguishable memory levels at sub-percent-class simulated variability~\cite{vourkas2021inmemory}; reaching imaging-relevant array scale and the endurance margin for continuous event processing (Milestone~N2) is the open device-level target. Array-level integration (Role~3: crossbar accelerators) offers the highest computational throughput, with 128$\times$64 HfO\textsubscript{2} arrays achieving 99.8\% yield and 5--8-bit precision~\cite{li2018analogue} and 2{,}048 analog conductance levels integrated on CMOS in a 256$\times$256 crossbar~\cite{rao2023thousands}; ADC/DAC peripheral co-design~\cite{shafiee2016isaac, xia2019memristive} is the open architectural target. Hybrid CMOS-memristor architectures balance power savings (75--79\% in simulation~\cite{yoon2024memristor}) against design complexity; the leading hybrid design awaits fabrication-level validation.

\textbf{RQ2: Material systems and device-level barriers.}
The reviewed material systems span metal oxides (HfO\textsubscript{2}, TaO\textsubscript{x}, SiN\textsubscript{x}), perovskites, molecular ferroelectrics, 2D materials (MoS\textsubscript{2}, h-BN), and biological polymers (silk protein). Key parameters are quantified in this review: D2D variability ranges from sub-percent-class simulated variability (SiN\textsubscript{x}~\cite{vourkas2021inmemory}) to 19.7\% (MoS\textsubscript{2}~\cite{tang2022waferscale}); endurance ranges from qualitative cyclic stability for perovskite dual-functional photomemristors~\cite{cao2023dual} to $>$10$^9$ polarization-reversal cycles for the perovskite ferroelectric class~\cite{wang2024perovskite}; and peripheral circuits (ADCs in particular) shape the area and energy budgets of memristive accelerators~\cite{shafiee2016isaac, xia2019memristive}. The cross-cutting materials challenge is meeting all six DVS-relevant requirements (low variability $<$5\%, high endurance $>$10$^{10}$ cycles, fast switching $<$10~\si{\micro\second}, retention $>$10 years, CMOS-compatible processing $<$400\si{\degreeCelsius}, and DVS-grade optical sensitivity for Role~1 candidates) on a single material system. Metal oxides currently lead on endurance and CMOS compatibility; perovskites lead on multi-mode functionality and offer the richest path to Role~1 integration.

\textbf{RQ3: Application maturity and deployment targets.}
Of 146 reviewed papers, ten operate at the memristor-DVS intersection at the device or array level, with the chip-prototype reference design as the most complete to date. AR/VR, medical imaging, and IoT remain \projected{} application domains awaiting their first demonstrators. Robotics, autonomous vehicles, and surveillance have device-level demonstrations (background subtraction~\cite{laiho2015memristive}, motion prediction~\cite{tan2023dynamic}, energy reduction~\cite{huang2025multimode}) that anchor the path to system-level validation with DVS event streams. The field sits at TRL~2--5; reaching TRL~6 requires a community memristor-DVS benchmark, endurance margins compatible with continuous event processing, and wafer-scale manufacturing uniformity at megapixel resolution.

The integration of memristors with dynamic vision sensors addresses a real and important problem: processing sparse, high-speed event streams at the sensor edge with minimal power. The field has produced a strong set of device-level demonstrations and one chip-prototype reference design; the next step is fabricated end-to-end measurement, and the milestones below define what that next step means in measurable terms.
Closing this gap means hitting the five engineering targets enumerated in Section~\ref{sec:gap-analysis}: end-to-end integration of a DVS sensor with a memristor compute layer, real-world deployment beyond benchtop demonstrations, evidence in every application domain, endurance margin for continuous event traffic, and a community-accepted memristor--DVS benchmark. Milestone~N1 ($\geq$90\% accuracy on DVS128-Gesture at $\leq$10\,mW system power and $\leq$5\,ms latency on a coupled DVS--memristor pipeline) is achievable with Role~3 integration and would be the field's first falsifiable end-to-end result; Milestone~M1 lifts the target to $\geq$640$\times$480 sensor resolution with a memristor backend at $\geq$85\% accuracy and $\leq$50\,mW. The roadmap is the path from the present demonstrations to those targets.

\subsection*{Threats to Validity}
\label{sec:threats-to-validity}
This review carries the standard limitations of a narrative literature assessment: coverage is bounded by three databases (IEEE Xplore, Web of Science, OpenAlex) and English-language publications, and venue-tier filtering under-represents work outside flagship venues. These constraints bound the precision of individual evidence grades but do not change the central finding.

\section*{Acknowledgments}
This work was supported in part by the U.S. National Science Foundation under Award No.~2432082.

\bibliographystyle{IEEEtran}
\bibliography{references}

\end{document}